\documentclass[10pt, journal, letterpaper, twocolumn]{IEEEtran}

\usepackage{amsfonts}
\usepackage[dvips]{graphicx}
\usepackage{times}
\usepackage{cite}
\usepackage{colortbl}
\usepackage{subfigure}
\usepackage{amsmath,amsthm}
\usepackage{array}
\usepackage{amssymb}

\newcommand{\bs}{\boldsymbol}

\usepackage{stfloats}
\usepackage{graphicx}
\usepackage{footnote}
\usepackage{booktabs}
\usepackage{array}
\usepackage[normalem]{ulem}
\usepackage[linesnumbered,lined,vlined,ruled,commentsnumbered]{algorithm2e}
\usepackage{algpseudocode}
\usepackage{subeqnarray}
\usepackage{cases}
\usepackage{threeparttable}
\usepackage{color}
\usepackage{url}
\usepackage{caption}
\usepackage{cancel}

\newtheorem{theorem}{Theorem}

\newtheorem{proposition}{Proposition}
\newtheorem{lemma}{Lemma}
\newtheorem{game}{Game}
\newtheorem{definition}{Definition}
\newtheorem{assumption}{Assumption}

\newtheorem{question}{Question}
\newtheorem{corollary}{Corollary}

\newtheorem{example}{Example}

\usepackage{mathtools}

\def\BibTeX{{\rm B\kern-.05em{\sc i\kern-.025em b}\kern-.08em
    T\kern-.1667em\lower.7ex\hbox{E}\kern-.125emX}}
    
\ifodd 0
\newcommand{\rev}[1]{{\color{blue}#1}}

\newcommand{\com}[1]{\textbf{\color{red} (Comment: #1) }}
\newcommand{\comg}[1]{\textbf{\color{blue} (COMMENT: #1)}}
\newcommand{\response}[1]{\textbf{\color{blue} (RESPONSE: #1)}}
\newcommand{\cross}[1]{{\color{red} \sout{#1}}}
\else
\newcommand{\rev}[1]{#1}

\newcommand{\com}[1]{}
\newcommand{\comg}[1]{}
\newcommand{\response}[1]{}
\newcommand{\cross}[1]{}
\fi

\begin{document}

\title{Pricing Fresh Data}
\author{
	Meng Zhang, \IEEEmembership{Member, IEEE}, 
	Ahmed Arafa, \IEEEmembership{Member, IEEE}, 
	Jianwei Huang, \IEEEmembership{Fellow, IEEE}, \\
	H. Vincent Poor, \IEEEmembership{Fellow, IEEE}\vspace{-10pt}\\
	
	\IEEEcompsocitemizethanks{ 
		\IEEEcompsocthanksitem
		Part of this work has been presented at WiOpt 2019 \cite{WiOpt}.
			\IEEEcompsocthanksitem
		M. Zhang is with the Department of Electrical and Computer Engineering, Northwestern University, IL, (e-mail: meng.zhang@northwestern.edu). A. Arafa is with the Department of Electrical and Computer Engineering,
University of North Carolina at Charlotte, NC (e-mail: aarafa@uncc.edu).
		J. Huang is with the Shenzhen Institute of Artificial Intelligence and Robotics for Society, the School of Science and Engineering, The Chinese University of Hong Kong, Shenzhen, Shenzhen 518172, China (corresponding author, e-mail: jianweihuang@cuhk.edu.cn)  H. V. Poor is with the Department of Electrical Engineering, Princeton University, NJ, (e-mail: poor@princeton.edu). 
		\IEEEcompsocthanksitem
		This work has been supported 
		in part by Shenzhen Institute of Artificial Intelligence and Robotics for Society, in part by the Presidential Fund from the Chinese University of Hong Kong, Shenzhen, and in part by the U.S. National Science Foundation  under Grants CCF-0939370 and CCF-1908308.
		}
	%
	
}
\maketitle
\thispagestyle{empty}

%
%
\vspace{-0.8cm}
\begin{abstract}
	We introduce the concept of {\it fresh data trading}, in which a destination user requests, and pays for, fresh data updates from a source provider, and
	data freshness is captured by the {\it age of information} (AoI) metric.
    Keeping data fresh relies on costly frequent data updates by the source, which motivates the source to {\it price fresh data}. In this work,
	the destination incurs an age-related cost, modeled as a general increasing function of the AoI. 
	The source 
	designs a pricing mechanism to maximize its profit, while the destination chooses a data update schedule to trade off its payments to the source and its age-related cost. 
	Depending on different real-time applications and scenarios, we study 
 both a {finite-horizon model and an infinite-horizon} model with time discounting.
The key challenge of designing the optimal pricing scheme lies in the destination's time-interdependent valuations, due to the nature of AoI, and 
the infinite-dimensional dynamic optimization.
To this end, we exploit three different dimensions in designing pricing by 
studying three pricing schemes: a {\it time-dependent} pricing scheme, in which the price for each update depends on when it is requested; a {\it quantity-based} pricing scheme, in which the price of each update depends on how many updates have been previously requested; and a simple {\it subscription-based} pricing scheme, in which the price per update is constant but the source charges an additional subscription fee.
Our analysis reveals that (1) the optimal subscription-based pricing maximizes the source's profit among all possible pricing schemes under both finite-horizon and infinite-horizon models; (2) the optimal quantity-based pricing scheme is only optimal with a finite horizon; and (3) the time-dependent pricing scheme, under the infinite-horizon model with significant time discounting, is asymptotically optimal. 
Numerical results show that the profit-maximizing pricing schemes can also lead to significant reductions in AoI and social costs, and that a moderate degree of time discounting is enough to achieve a close-to-optimal time-dependent pricing scheme.

\end{abstract}

\section{Introduction}
\subsection{Motivations}
\setcounter{page}{0}
 Information usually has the greatest value when it is fresh \cite[p. 56]{fresh}. 
 Data freshness
 is becoming increasingly significant due to the fast growth of the number of mobile devices and the dramatic increase of real-time systems. For instance, real-time knowledge of traffic information and the speed of motor vehicles is crucial in autonomous driving and unmanned aerial vehicles. Hence, it has driven the new metric to measure data freshness, namely {\it age-of-information} (AoI) introduced in \cite{AoI2}, defined as the time elapsed since the freshest data has reached its destination.
Real-time systems range from
Internet-of-Things (IoT) industry, multimedia, cloud-computing services, real-time data analytics, to even financial markets.
More specifically, examples of real-time applications demanding timely data updates include monitoring, data analytics and control systems, phasor data updates in power grid stabilization systems;
  examples of real-time datasets include real-time map and traffic data, e.g., the Google Maps Platform \cite{map}. The systems involving these applications and datasets put high emphasis on the data freshness.

\begin{figure}
	\begin{centering}
		\includegraphics[scale=0.45]{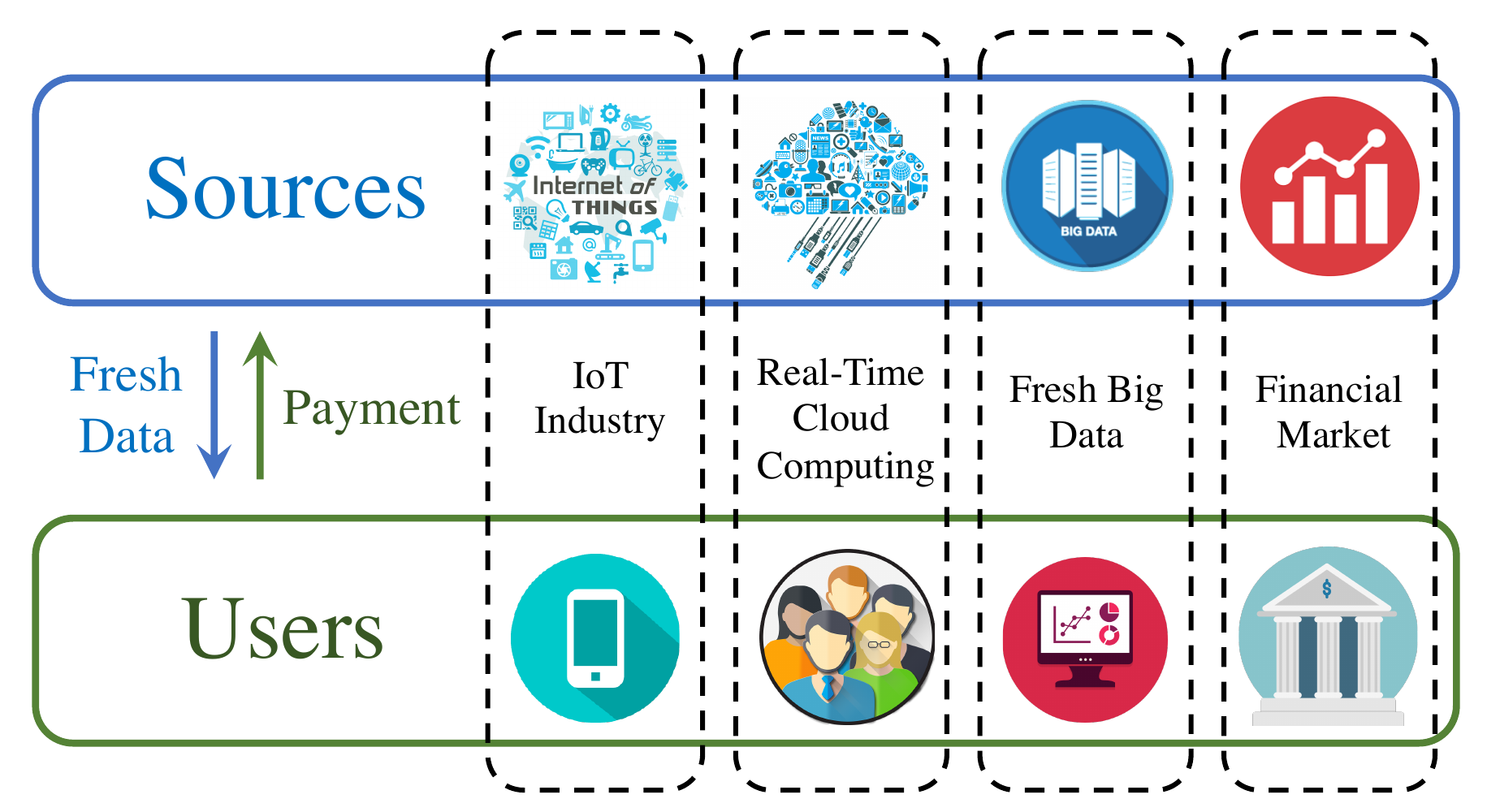}
		\vspace{-0.15cm}
		\caption{\rev{Examples of potential fresh data markets.}}
		\label{System}
		\vspace{-0.4cm}
	\end{centering}
\end{figure}

Despite the increasing significance of fresh data, 
keeping data fresh relies on frequent data generation, processing, and transmission, which can lead to significant operational costs for the data sources (providers).
Such operational costs make \textit{pricing design} of an essential role in the fresh data trading interaction between data sources and data destinations (users), as pricing provides an incentive for the sources to update the data and prohibits the destinations from requesting data updates unnecessarily often. Furthermore, in addition to enabling necessary fresh data trading, pricing design is also one of the core techniques of \textit{revenue management}, facilitating data sources' profit maximization.
 
 The pricing for fresh data, however, is under-explored, as all existing pricing schemes for communication systems serve to control the network congestion level and assume that a consumer's satisfaction with the service depends mainly on the quantity/quality of the service received without considering its timeliness.  
Fig. \ref{System} illustrates the interaction in  fresh data markets between data providers and users requesting fresh data. 
This paper studies the specifics of fresh data trading with a single source-destination pair, aiming at answering the following question:
\begin{question}
	How should the source choose a pricing scheme to maximize its profit in fresh data trading?
\end{question}

\subsection{Approaches and Challenges}
Motivated by different types of fresh data businesses, we consider both a \textit{predictable-deadline} model and an \textit{unpredictable-deadline} model. In the former case, the source and the destination may only interact sequentially for a (potentially short-term) finite horizon (e.g., deadline-aware cloud computing tasks \cite{cloud}). In the latter case, such 
an interaction is relatively long-term such that the destination and the source may not know the exact deadline of such a  fresh data business (e.g., uncertain completion-time cloud computing tasks \cite{cloud2}). 


For both the predictable-deadline model and the unpredictable-deadline model, we 
study three types of pricing schemes \rev{by exploiting three different dimensions,
 namely \textit{time, quantity, and subscription}}, described in the following:
 \begin{itemize}
     \item \textit{Time-dependent pricing scheme}: The source of fresh data  prices each data update based on the time at which the update is requested. Due to the nature of the AoI, the destination's desire for updates increases as time (since the most recent update) goes by, which makes it natural to explore this time sensitivity. This pricing scheme is also motivated by practical pricing schemes for mobile networks (in which users are not age-sensitive) \cite{Pricing3,time2,time3,time4}.
     \item \textit{Quantity-based pricing scheme}: The price for each update depends on the number of updates requested so far (but does not depend on the timing of the updates) \cite{Price}. 
The source may reward the destination by reduced prices for each additional request
to incentivize more fresh data updates.
Such a pricing scheme is motivated by practical pricing schemes for data services (e.g., for data analytics services \cite{api} and cloud computing services \cite{map,RimuHosting}). For instance, the storage provider RimuHosting charges a smaller price for each additional gigabyte of storage purchased \cite{RimuHosting}.
\item \textit{Subscription-based pricing scheme}: 
The source charges a one-time subscription price and a 
flat-rate usage price for each update (instead of differentiating the price over time or quantity dimensions). Such a pricing scheme is motivated by practical pricing schemes for  mobile network data plans and services \cite{Pricing3,tariff} and enjoys a low implementation complexity as it is characterized by two parameters only.

 \end{itemize}
 

\begin{table*}[t]
\centering
\begin{tabular}{ccc}
\hline
                           & \multicolumn{2}{c}{Profit maximization (among all possible pricing schemes)}          \\ \hline
 \rowcolor[gray]{.95}                          & Finite-horizon model & Infinite-horizon model \\ \hline \hline
Time-Dependent Pricing     & $\times$                   & Asymptotically optimal       \\ \hline
 \rowcolor[gray]{.95}   Quantity-Based Pricing     & $\checkmark$               & $\times$                     \\ \hline
Subscription-Based Pricing & $\checkmark$               & $\checkmark$                 \\ \hline
\end{tabular}
\vspace{-0.1cm}
\caption{Summary of key results.}\label{Tab1}
\vspace{-0.3cm}
\end{table*}

Our goal is to explore these three different aforementioned pricing schemes and address the following question:
\begin{question}
\rev{How profitable it is to exploit the time, quantity, and subscription dimensions in the pricing design of fresh data?}
\end{question}

{The nature of data freshness poses the threefold challenge of designing the above pricing schemes. 
First, the destination's valuation is \textit{time-interdependent}, which makes it significantly different from conventional (physical or digital) goods
(e.g., \cite{Pricing3,time2,time3,time4}).
 That is, the desire for an update at each time instance depends on the time elapsed since the latest update. Hence, the source's pricing scheme choice needs to take such interdependence into consideration.  Second, the flexibility in different pricing choices 
renders the optimization over (infinitely) many dimensions. 
Third, the time discounting infinite-horizon model constitutes a challenging continuous-time dynamic programming problem.}

The key results and contributions of this paper are summarized as follows:
\begin{itemize}
	\item \emph{Fresh Data Trading Modeling with General AoI Cost. } To the best of our knowledge, this paper presents the first study of the source pricing scheme design in fresh data trading, in which we consider
	a general increasing age-related cost function for the destination.

    \item \emph{Profit Maximizing Pricing.} Under the finite-horizon model, our analysis reveals that exploiting the quantity dimension or the subscription dimension alone can maximize the source's profit. On the other hand, under the infinite-horizon model, only the subscription-based pricing can achieve profit maximization.


    \item \emph{Effectiveness of Exploiting the Time Dimension.} We show that profitability of exploiting the time dimension depends on both the deadline type and the \textit{time discounting}. In particular, the optimal time-dependent pricing can be  \textit{time-invariant} under the finite-horizon model, and hence renders exploitation of the time dimension ineffective. On the other hand, under the infinite-horizon model with significant time discounting, time-dependent pricing asymptotically maximizes the source's profit among all possible pricing schemes.

\item \emph{Numerical Results.} Our numerical studies show that the quantity-based pricing scheme and the subscription-based pricing may also lead to significant reductions in AoI and social costs, incurring up to $41\%$ of less AoI and
up to $54\%$ less social cost, compared against the optimal time-dependent pricing scheme. 
In addition,  we show that
the  time-dependent  pricing can be asymptotically profit-maximizing even under moderate time discounting.

\end{itemize}

\rev{Table \ref{Tab1} summarizes the key results regarding the three pricing schemes analyzed in this paper.}

We organize the rest of this paper as follows. In Section \ref{Relate}, we discuss some related work. In Section \ref{Sysm}, we describe the system model and the game-theoretic problem formulation. In Sections \ref{Predict} and \ref{Unpredict}, we develop the time-dependent, the quantity-based and subscription-based pricing schemes under the finite-horizon model and the infinite-horizon model, respectively. We provide some numerical results in Section \ref{Numerical} to evaluate the performance of the three pricing schemes, and conclude the paper in Section \ref{Conclusion}.

\section{Related Work}\label{Relate}
In recent years, there have been many excellent works focusing on the optimization of scheduling policies that minimize the AoI in various system settings, e.g., \cite{AoI2,AoI3,AoI4,AoI42,AoI5,AoI6,AoI77,New1,New2,energy1,AoI7,AoI12,AoI13}. In \cite{AoI2}, Kaul \textit{et al.} recognized the importance of real-time status updates in networks. In \cite{AoI4,AoI42}, He \textit{et al.} investigated the
NP-hardness of minimizing the AoI in  scheduling general wireless networks.
In \cite{AoI5}, Kadota \textit{et al.} studied the scheduling problem in a wireless network with a single base station and multiple destinations. In \cite{AoI6}, Kam \textit{et al.} investigated the AoI for a status updating system through a network cloud. In \cite{AoI7}, Sun \textit{et al.} studied the optimal management of the fresh information updates. In \cite{AoI77}, Bedewy \textit{et al.} studied a joint sampling and transmission scheduling problem in a multi-source system.
References \cite{New1} and \cite{New2} studied the optimal wireless network scheduling with an interference constraint and a throughput constraint, respectively.
The AoI consideration has also gained some attention in energy harvesting communication systems, e.g., \cite{AoI3,energy1,AoI12,AoI13}, and Internet of Things systems, e.g., \cite{NewAge1,NewAge2}.
Several existing studies focused on game-theoretic interactions in interference channels, e.g., \cite{AoI10,AoI11}.
\rev{All the aforementioned works have not considered the economic interactions among sources and destinations.}

\rev{More related AoI studies are those pertaining to the economics of fresh data and information \cite{AoIEcon1,AoIEcon2,AoIEcon3}.} In \cite{AoIEcon1}, a repeated game is studied between two AoI-aware platforms, yet without studying pricing schemes. References \cite{AoIEcon2,AoIEcon3} considered timely systems in which the destinations design pricing schemes to incentivize sensors to provide fresh updates. Different from \cite{AoIEcon2,AoIEcon3}, our considered pricing schemes are designed by the source, which is motivated by most practical communication/data systems in which sources are price designers while the destinations are \textit{myopic} instead of \textit{forward-looking} as we consider in this work.

\section{System Model}\label{Sysm}
In this section, we introduce the system model of a single-source single-destination information update system and formulate the corresponding pricing scheme design problem.

\subsection{System Overview}

\subsubsection{Single-Source Single-Destination System} 
We consider an information update system, in which one source node generates data packets  and sends them to one destination through a channel. For instance, {Amazon Web Services (the source) provides real-time data processing and analytics services to deliver client-specific data for each individual client (the destination), e.g., Airbnb  \cite{AWS}.}


{We note that the single-source single-destination model has been widely considered in the AoI literature (e.g.,
	\cite{AoI3,AoI6,AoI7,AoI12,AoI13}). The insights (such as the potential optimal pricing structures) derived from this model allow extensions to multi-destination scenarios.}\footnote{{The system constraints (e.g.,
	congestion and interference constraints) in a multi-destination model can make the joint scheduling and pricing scheme design much more challenging, as it involves competition among destinations and requires more sophisticated game-theoretic analysis.}}


\subsubsection{Data Updates  and Age-of-Information}  

We consider a fixed time period of $\mathcal{T}=[0,T]$, during which the source sends its updates to the destination. We consider a generate-at-will model (as in, e.g., \cite{AoI3,energy1,AoI12,AoI13}), in which the source is able to generate and send a new update when requested by the destination. Updates reach the destination instantly, with negligible transmission time (as in, e.g., \cite{energy1,AoI12}).\footnote{ This assumption is practical when inter-update times are on a scale that is order of magnitudes larger than the transmission times of the updates themselves. }

We denote by $S_k\in\mathcal{T}$ the transmission time of the $k$-th update. The set of all update time instances is $\mathcal{S}\triangleq\{S_k\}_{1 \leq k\leq K}$, where $K$ is the number of total updates, i.e., $|\mathcal{S}|=K$ with $|\cdot|$ denoting the cardinality of a set. The set $\mathcal{S}$ (and hence the value of $K$) is
the destination's decision. {We use $\Phi$ to denote the feasible set of  $\mathcal{S}$ satisfying $S_k\geq S_{k-1}$ for all $1\leq k\leq K$.}
Let $x_{k}$ denote the $k$-th update interarrival time, which is the time elapsed between the generation of ($k-1$)-th update and  $k$-th update, i.e., $x_{k}$ is\footnote{We read $S_0$ as $0$ and $S_{K+1}$ as $T$.}
\begin{align}
x_{k}\triangleq S_{k}-S_{k-1},~\forall k\in\mathcal{K}(K+1),\label{x}
\end{align}
where {$\mathcal{K}(K)\triangleq\{1,...,K\}$.}
Let $\bs{x}\triangleq \{x_{k}\}_{k\in\mathcal{K}(K+1)}$ be the vector of update interarrival times.\footnote{Throughout this paper, we use $(\bs{x},K)$ and $\mathcal{S}$ to denote the update policy interchangeably.}

The following definition characterizes the freshness of data:
\vspace{-0.25cm}
\begin{definition}[Age-of-Information (AoI)]
The age-of-information $\Delta_t(\mathcal{S})$ at time $t$ is \cite{AoI2}
\begin{align}
\Delta_{t}(\mathcal{S})=t-U_{t},\label{AoI}
\end{align}
where $U_t$ is the time stamp of the most recently received update before time $t$, i.e., $U_{t}=\max_{S_{k}\leq t}\{S_{k}\}$.
\end{definition}

\begin{figure}
	\begin{centering}
		\includegraphics[scale=0.375]{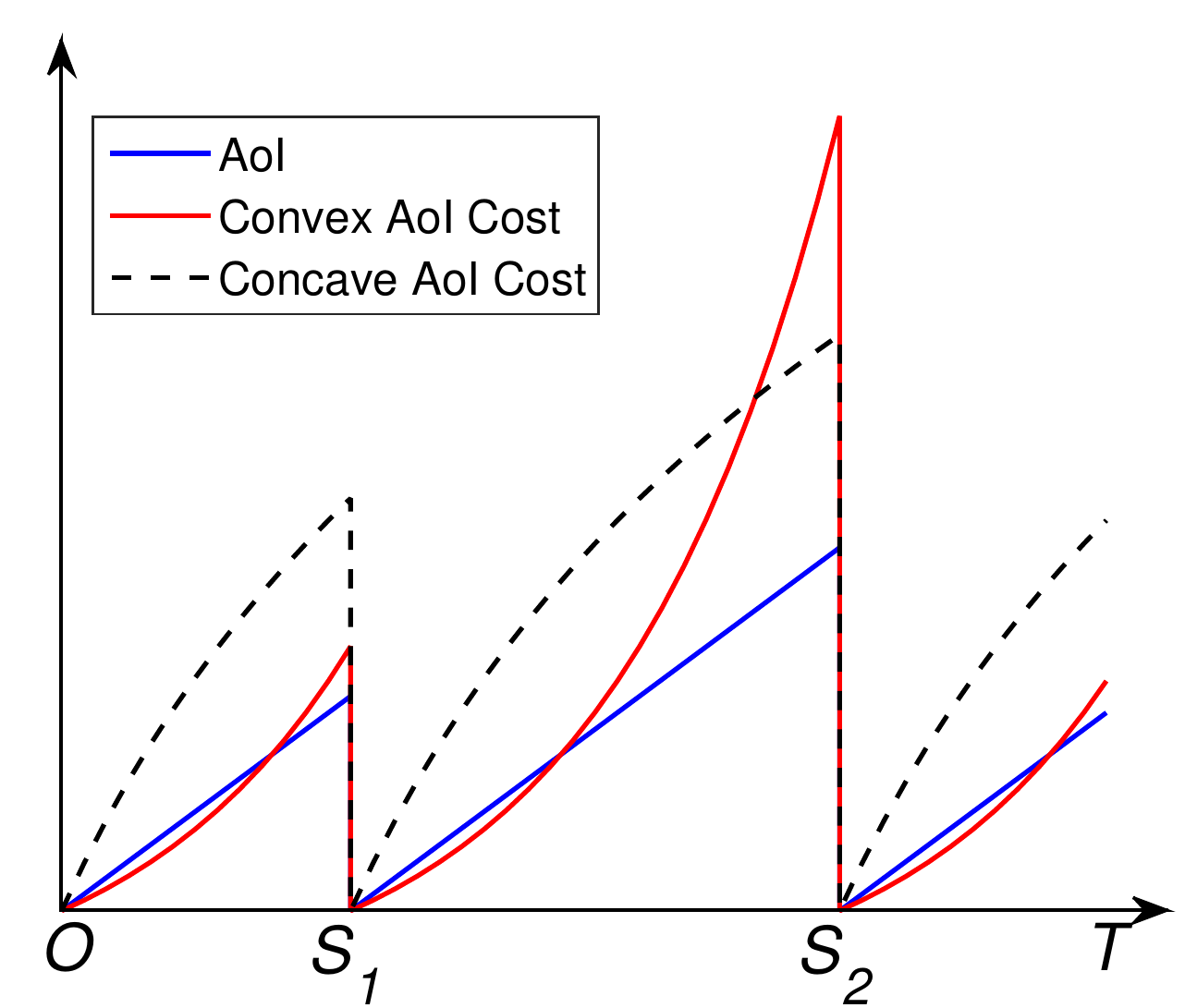}
	\vspace{-0.1cm}
		\caption{Illustrations of AoI $\Delta_t$ and two types of AoI costs $f(\Delta_t)$. There are two updates at $S_1$ and $S_2$.}
			\vspace{-0.4cm}
		\label{AoI}
	\end{centering}
\end{figure}

\subsubsection{Destination's General AoI Cost} 

The destination experiences an AoI cost $f(\Delta_t)$ related to its desire for the new data update (or dissatisfaction of stale data).
We assume that $f(\Delta_t)$ is a general \textit{increasing} function in $\Delta_t$. For instance, \rev{a convex AoI cost implies the destination gets more desperate when its data grows stale,} an example of which
is $f(\Delta_t)=\Delta_t^\kappa$ for $\kappa\geq 1$, which exists in the online learning in real-time applications such as online advertisement placement and online Web ranking \cite{online1,online2}. 
Fig. \ref{AoI} illustrates  the AoI, a convex AoI cost function and a concave AoI cost function. {We next introduce the following AoI-related notations:
\vspace{-0.3cm}
\begin{definition}[Aggregate and Cumulative AoI Cost]\label{CAoI}
The destination's aggregate AoI cost $\Gamma(\mathcal{S})$  and the cumulative AoI Cost $F(x)$ for each interarrival time $x$ (between two updates) are
\begin{align}
\Gamma(\mathcal{S})\triangleq\int_{0}^{T} f(\Delta_t(\mathcal{S}))dt \quad {\rm and}\quad F(x)\triangleq\int_{0}^x f(t)dt.\label{gamma}
\end{align}
\end{definition}
Based on Definition \ref{CAoI}, we have $\Gamma(\mathcal{S})=\sum_{k=1}^{K+1}F(x_k)$.}

\subsubsection{Source's Operational Cost and Pricing} Let $\bar{x}=T/(K+1)$ be the  average interarrival time.
We use $c\left({\bar{x}}\right)$ to denote the source's \textit{operational cost per update}, which is modeled as a \textit{non-increasing} and \textit{convex} function.\footnote{Non-increasingness indicates that the cost per update can only decrease when the source updates less frequently, and convexity implies that the incremental reduction in the cost per update decreases in the average interarrival time.}
This can represent sampling costs in case the source is an IoT service provider, the computing resource consumption in case the source is a cloud computing service provider\footnote{In particular, the non-increasing and convex average operational cost satisfies the sublinear speedup: the consumed computing resources multiplied by the completion time) for each task is increasingly higher under a shorter completion time  \cite{CPU2}.}, and 
transmission costs in case the source is a network operator\footnote{By the Shannon–Hartley theorem, the consumed energy per achievable bit is decreasing and convex in transmission time.}. 
Such an operational cost generalizes the fixed sampling cost model in \cite{AoIEcon2}. We have the following definition for operational cost:
\vspace{-0.3cm}
{\begin{definition}[Operational Cost]
The source's operational cost $C(K)$ is given by
\begin{align}
{C(K)\triangleq K\cdot c\left(T/(K+1)\right)}. \label{OC}
\end{align}
\end{definition}
As \eqref{OC} indicates, update policies leading to the same $K$ incur the same operational cost for the source.
Since $c(\cdot)$ is non-increasing and convex, $C(K)$ is increasing and convex in $K$. }

The source designs the pricing scheme, denoted by $\Pi$, for sending the data updates.
A  pricing scheme may exploit three dimensions: \textit{time, quantity,} and \textit{subscription}. Specifically, we consider a {\it time-dependent} pricing scheme $\Pi_t$, in which the price for each update depends on $t$, i.e., when it is requested; a {\it quantity-based} pricing scheme $\Pi_q$, in which the price for each update varies; and a {\it subscription-based} pricing scheme $\Pi_s$, in which the source charges an additional subscription fee.\footnote{As mentioned, these pricing schemes are motivated by (i) the time-sensitive  demand for an update due to the nature of AoI, and (ii) the wide consideration  of time-dependent, quantity-based, and subscription-based pricing schemes in practice \cite{Pricing3,tariff}.} We next define the destination's total payment $P(\mathcal{S},\Pi)$, which depends on the destination's update policy $\mathcal{S}$ and the source's pricing scheme $\Pi$ to be specified in Section \ref{Predict}. 


\subsection{Stackelberg Games}

\begin{figure}[t]
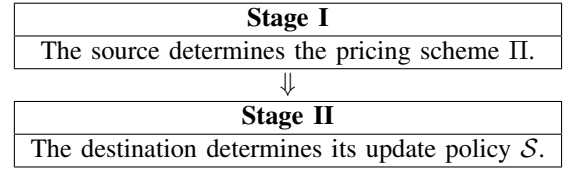

	\centering
	\begin{tabular}{c}
		\hline
		\multicolumn{1}{|c|}{ \textbf{Stage I}}        \\ \hline
		\multicolumn{1}{|c|}{The source determines the pricing scheme $\Pi$.} \\ \hline
		$\Downarrow   $                                  \\ \hline
		\multicolumn{1}{|c|}{\textbf{Stage II}}        \\ \hline
		\multicolumn{1}{|c|}{The destination determines its update policy $\mathcal{S}$.}         \\ \hline
	\end{tabular}
	
	\caption{Two-stage Stackelberg game.}	\label{Stack1}
	\vspace{-0.4cm}
\end{figure}
We model the interaction between the source and the destination as a
two-stage Stackelberg game, as shown in Fig. \ref{Stack1}. Depending on different applications and the associated business, we categorize the interactions between the source and the destination into a \textit{finite-horizon model} and an \textit{infinite-horizon model}. In the former case, the interaction take place for a (potentially short-term) finite horizon (e.g., deadline-aware cloud computing tasks \cite{cloud}). In the latter case, the interaction is longer-term such that the destination and the source may not know the exact deadline (e.g., uncertain completion-time cloud computing tasks \cite{cloud2}). 

{Given the aggregate AoI cost in \eqref{gamma}, a feasible pricing scheme $\Pi$ needs to satisfy an \textit{individual rationality} constraint: the destination should be no worse off than receiving no update; otherwise, the pricing scheme drives away  the destination.
Let $\mathcal{S}^{*}(\Pi)$ be the destination's optimal update policy in response to the pricing scheme $\Pi$ chosen by the source, which will be defined soon. Based on this, any pricing scheme $\Pi$ needs to satisfy the individual rationality constraint:
\begin{align}
    \Gamma(\mathcal{S}^{*}(\Pi))+P(\mathcal{S}^{*}(\Pi),\Pi)\leq F(T). \label{IRP}
\end{align}
That is, the destination should achieve an overall cost no larger than a no-update policy $F(T)$.
The following definition of the Stackelberg Game captures the interaction between the source and the destination:
\begin{game}[Source-Destination Interaction Game] The interaction between the source and the destination involves two stages:
\begin{itemize}
    \item  In Stage I, the source decides on the pricing scheme $\Pi$ at the beginning of the period, in order to maximize its profit, given by:
\begin{subequations}\label{source-1}
	\begin{align}
&{\rm \mathbf{Source-F:}}\nonumber\\
~&\max_{\Pi}~P(\mathcal{S}^{*}(\Pi),\Pi)-C(|\mathcal{S}^{*}(\Pi)|)\\
	&~~{\rm s.t.}~~\Pi\in\{\Pi: \eqref{IRP}, \pi, p_k(t)\geq 0,~\forall t\in\mathcal{T}, k\in\mathbb{N}\}.
	\end{align}
\end{subequations}
\item In Stage II, given the source's decided pricing scheme $\Pi$, the destination decides on its update policy 	to minimize its overall cost (aggregate AoI cost plus payment):
\begin{align}\label{destination-1}
{\rm \mathbf{Destination-F:}} ~\mathcal{S}^{*}(\Pi)\triangleq\arg\min_{\mathcal{S}\in\Phi}~\Gamma(\mathcal{S})+P(\mathcal{S},\Pi).
\end{align}
\end{itemize}
\end{game}}

We will analyze 
the pricing scheme design problems in Section \ref{Predict}.
{In Section \ref{Unpredict}, we will specify and analyze a new game based on an infinite-horizon model with time discounting. }


%
%


\section{Finite-Horizon Model}\label{Predict}

In this section, we will first derive the upper bound of the source's achievable profit when there is a finite deadline $T$. We will then separately consider three special cases of the pricing $\Pi$ by exploiting different dimensions: time-dependent pricing $\Pi_t$, quantity-based pricing $\Pi_q$, and subscription-based pricing $\Pi_s$. We will show the existence of the optimal $\Pi_t$ and $\Pi_q$ schemes that can maximize the source's profit among all possible pricing schemes.



\subsection{Social Cost Minimization and Surplus Extraction}
To evaluate the performances of the pricing schemes to be studied, we first consider an achievable upper bound of the source's profit for any pricing scheme
in this subsection. 
Note that the outcome attaining such an upper bound of the profit collides with the achievement of another system-level goal, namely the \textit{social optimum}:
\vspace{-0.2cm}
\begin{definition}[Social Optimum]
A social optimum update policy $\mathcal{S}^o$ solves the following social cost minimization problem:
\begin{align}
{{\rm \mathbf{SCM-F:}}~\mathcal{S}^o\triangleq\arg\min_{\mathcal{S}\in\Phi}~C(|\mathcal{S}|)+\Gamma(\mathcal{S}).\label{SCM-F}}
\end{align}
\end{definition}
That is, the socially optimal update policy minimizes the source's operational cost \rev{$C(|\mathcal{S}|)$} and the destination's AoI cost \rev{$\Gamma(\mathcal{S})$} combined. We further introduce the following definition:
{\begin{definition}[Surplus Extraction]\label{Def7}
A pricing scheme $\Pi$ is surplus-extracting if it satisfies
\begin{align}
P(\mathcal{S}^{*}(\Pi),\Pi)=F(T)-\Gamma(\mathcal{S}^{*}(\Pi))~{\rm and}~
    \mathcal{S}^{*}(\Pi)=\mathcal{S}^o,
\end{align}
where $\mathcal{S}^{*}(\Pi)$ and $\mathcal{S}^o$ are defined in \eqref{destination-1} and \eqref{SCM-F}, respectively.
\end{definition}}
That is, the surplus extracting pricing leads to a payment equal to the destination's overall AoI cost reduction, i.e., the overall AoI cost with no updates \rev{$F(T)$} minus the \rev{overall AoI cost under a socially optimal update policy $\Gamma(\mathcal{S}^o)$}. We are now ready to show that the optimality of a surplus-extracting pricing:
\begin{lemma}\label{LLL1}
Under {the finite-horizon model}, {every} surplus-extracting pricing scheme (satisfying Definition \ref{CAoI}) maximizes the source's profit  among all possible pricing schemes, i.e., it corresponds to the optimal solution of the problem in \eqref{source-1}.
\end{lemma}

We prove Lemma \ref{LLL1} in Appendix \ref{ProofLLL1}.
In later analysis, we will show that the optimal quantity-based pricing and the optimal subscription-based pricing schemes are surplus-extracting for the finite-horizon case. However, the time-dependent pricing in general is not.

\subsection{Time-Dependent Pricing Scheme}
We first consider a (pure) \textit{time-dependent} pricing scheme $\Pi_t=\{p(t)\}_{t\in\mathcal{T}}$, in which the price $p(t)$ for each update  depends on the time at which \rev{each update $k$ is requested (i.e., $S_k$)} and does not depend on the number of updates so far. Hence, the payment is $P(\mathcal{S},\Pi_t)=\sum_{k=1}^{K}p(S_k)$.

We derive the \textit{(Stackelberg subgame perfect)} equilibrium price-update profile  $(\Pi^{\rm *}_t,\mathcal{S}^{\rm *}(\Pi^{\rm *}_t))$ by  backward induction. First,  given any pricing scheme $\Pi_t$  in Stage I, we characterize the destination's update policy $\mathcal{S}^{\rm *}(\Pi_t)$ that minimizes its overall cost in Stage II. Then in Stage I, by characterizing the equilibrium pricing structure, we convert the continuous function optimization into a vector one, based on which
we characterize the source's optimal pricing scheme $\Pi_t^{\rm *}$.

\subsubsection{Destination's Update Policy in Stage II} 
We analyze the destination's update policy under arbitrary $\Pi_t$ within the fixed time period $[0,T]$.
Recall that  $K$ is  the total number of updates and $x_k$ defined in \eqref{x} is the $k$-th interarrival  time.
Given the pricing scheme $\Pi_t$, we can simplify the destination's overall cost minimization problem in \eqref{destination-1} as
\vspace{-0.3cm}
\begin{subequations}
	\begin{align}
	&\min_{K\in\mathbb{N}\cup\{0\},\bs{x}\in\mathbb{R}_{++}^{K+1}}~\sum_{k=1}^{K+1}F(x_k)+\sum_{k=1}^Kp\left(\sum_{j\leq k}x_j\right),\\
	&~~~~~~~~{\rm s.t.}~~~~~~~~\sum_{k=1}^{K+1}x_k=T,
	\end{align}
\end{subequations}
where $\mathbb{R}_{++}^{K}$ is the space of $(K)$-dimensional positive vectors (i.e., the value of every entry is positive).


%

%

\begin{figure}
	\begin{centering}
		\includegraphics[scale=0.23]{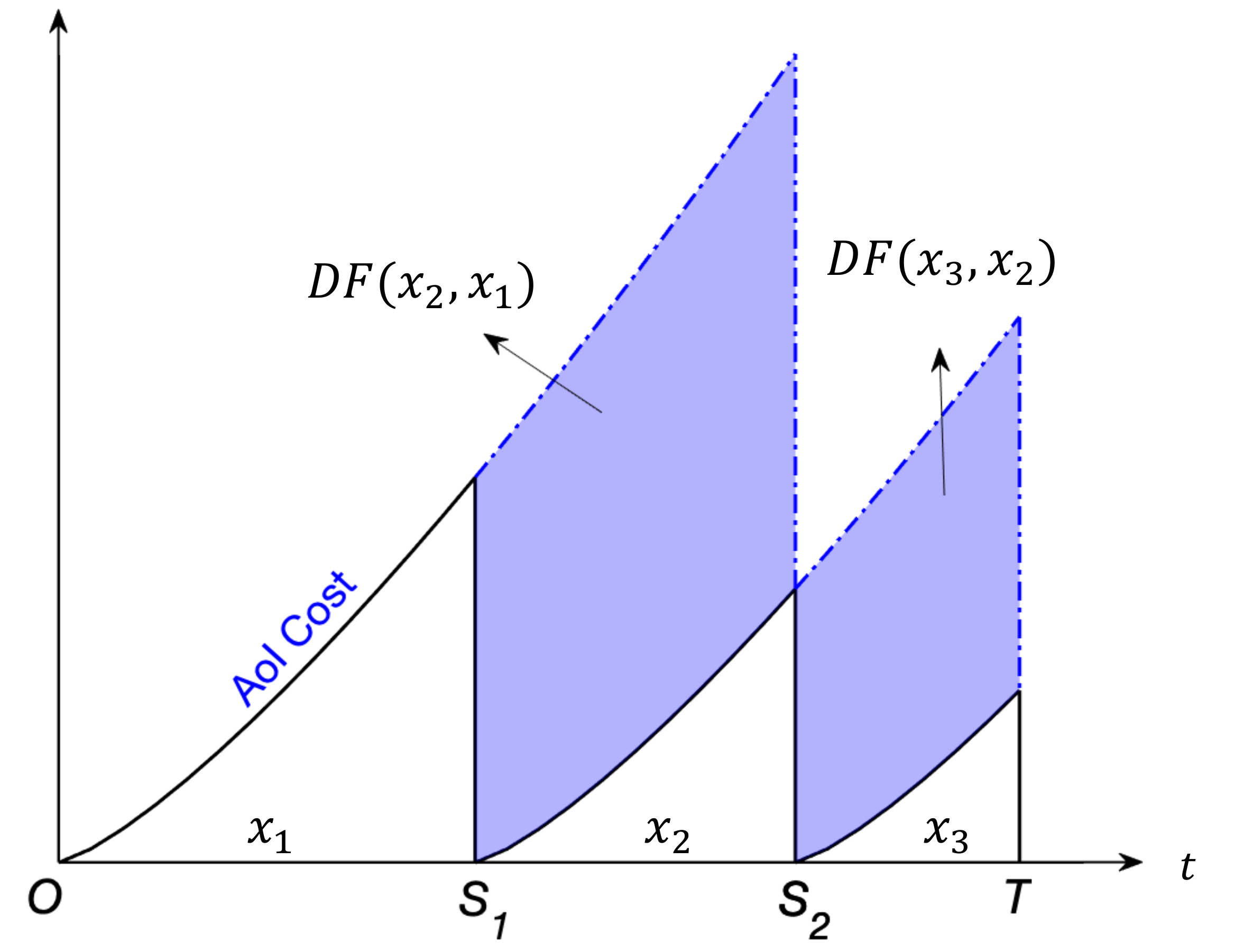}
		\vspace{-0.1cm}
		\caption{An illustrative example of the differential aggregate AoI cost function and Lemma \ref{L1}. }
		\label{Lemma2}
	\end{centering}
	\vspace{-0.4cm}
\end{figure}

To understand how \rev{the destination evaluates fresh data,} we introduce the following definition:
\begin{definition}[Differential Aggregate AoI Cost]
The differential aggregate AoI cost function is
\begin{align}
DF(x,y)\triangleq \int_0^x [f(t+y)-f(t)]dt.\label{DF}
\end{align}
\end{definition}
As illustrated in  Fig. \ref{Lemma2}, for each update $k$,  $DF(x_{k+1},x_k)$ is the aggregate AoI cost increase if the destination changes its update policy from $\mathcal{S}$ to  $\mathcal{S}\backslash\{S_k\}$ (i.e., removing the update at  $S_k$).
\rev{We now derive the optimal time-dependent pricing based on \eqref{DF} in the following lemma:}
\vspace{-0.25cm}
\begin{lemma}\label{L1}
	Any equilibrium price-update tuple $(\Pi_t^{*},K^{\rm *, T},\bs{x}^{\rm *, T})$  should satisfy\footnote{We use  $(K^{\rm *, T},\bs{x}^{\rm *, T})$ to denote the equilibrium update policy under the optimal time-dependent pricing, i.e., $(K^{\rm *, T},\bs{x}^{\rm *, T})=(K^{\rm *}(\Pi_t^{*}),\bs{x}^{\rm *}(\Pi_t^{*}))$.} 
	\begin{align}
	p^{\rm *}\left(\sum_{j=1}^kx_j^{\rm *, T}\right)=
	DF(x_{k+1}^{\rm *, T},x_k^{\rm *, T}),~&~\forall k\in\mathcal{K}(K^{\rm *, T}+1).\label{price}
	\end{align}
\end{lemma}
We present the proof of Lemma \ref{L1} in Appendix \ref{ProofL1}.
Intuitively, the differential aggregate AoI cost equals  the destination's \textit{maximal willingness to pay} for each update.
Note that given that the optimal time-dependent pricing scheme satisfies \eqref{price}, there might exist multiple optimal update policies as the
solutions of problem \eqref{destination-1}. This may lead to a multi-valued source's profit and thus an ill-defined  problem \eqref{source-1}. {To ensure the uniqueness of the received profit for the source \rev{without affecting the optimality to the source's pricing problem}, 
one can impose infinitely large prices to
ensure that  the destination does not update at any time instance other than $\sum_{j=1}^kx_j^{\rm *, T}$ for each $k\in\mathcal{K}(K^{\rm *, T}+1)$. Together with the pricing in Lemma \ref{L1}, it leads to a unique update policy.}

\subsubsection{Source's Time-Dependent Pricing Design in Stage I}
Based on Lemma \ref{L1},  we can reformulate the time-dependent pricing scheme as follows. In particular, the decision variables in problem \eqref{Refor} correspond to the interarrival time interval vector $\bs{x}$ instead of the continuous-time pricing function $p(t)$. 
By converting a functional optimization problem into a finite-dimensional vector optimization problem, we simplify the problem as follows.
\begin{proposition}\label{P1}
	The time-dependent pricing problem  in \eqref{source-1} is equivalent to the following problem:
	\begin{subequations}\label{Refor}
		\begin{align}
		\max_{K\in\mathbb{N}\cup\{0\},\bs{x}\in\mathbb{R}^{K+1}_{++}}&~~\sum_{k=1}^{K}DF(x_{k+1},x_k)-C(K),\label{Refor1}\\
		{\rm s.t.}~~~~~~~&~~\sum_{k=1}^{K+1}x_k=T\label{dsa}.
		\end{align}
	\end{subequations}
\end{proposition}

We prove Proposition \ref{P1} in Appendix \ref{ProofP1}.
Note that  \rev{the constraint in \eqref{IRP} is automatically satisfied here}, as the destination can always choose a no-update policy (i.e., $K=0$) leading to a cost of $F(T)$ under any $\Pi_t$.
To rule out trivial cases with no update at the equilibrium, we adopt the following assumption throughout this paper:
	\begin{assumption}\label{A2}
		The source's operational cost function $C(K)$ satisfies $C(1)\leq DF(T/2,T/2).$
	\end{assumption}
	{Assumption \ref{A2} ensures that the operational cost for one update $C(1)$ is not larger than the source's willingness to pay such an update.}
We consider the convex AoI function to derive some insightful results:
\begin{proposition}\label{P222}
When Assumption \ref{A2} holds and the AoI function $f(x)$ is convex, then there will be only one update (i.e., $K^{\rm *, T}=1$) under any equilibrium time-dependent pricing scheme. 
\end{proposition}
\vspace{-0.25cm}
{The intuition behind Proposition 2 is that a convex AoI cost leads to an accelerated increase in the destination's willingness to pay as AoI increases. Hence, it is most profitable to charge a relatively high price to induce only one update.}
We can prove Proposition \ref{P222}  by induction, showing that for an arbitrary time-dependent pricing scheme yielding more than $K>1$ updates ($K$-update pricing), there  always exists a pricing scheme with a single-update equilibrium that is more profitable.	Based on the above technique,  we can show that the above argument works for any increasing convex AoI cost function. 

\begin{figure}
	\begin{centering}
		\includegraphics[scale=0.35]{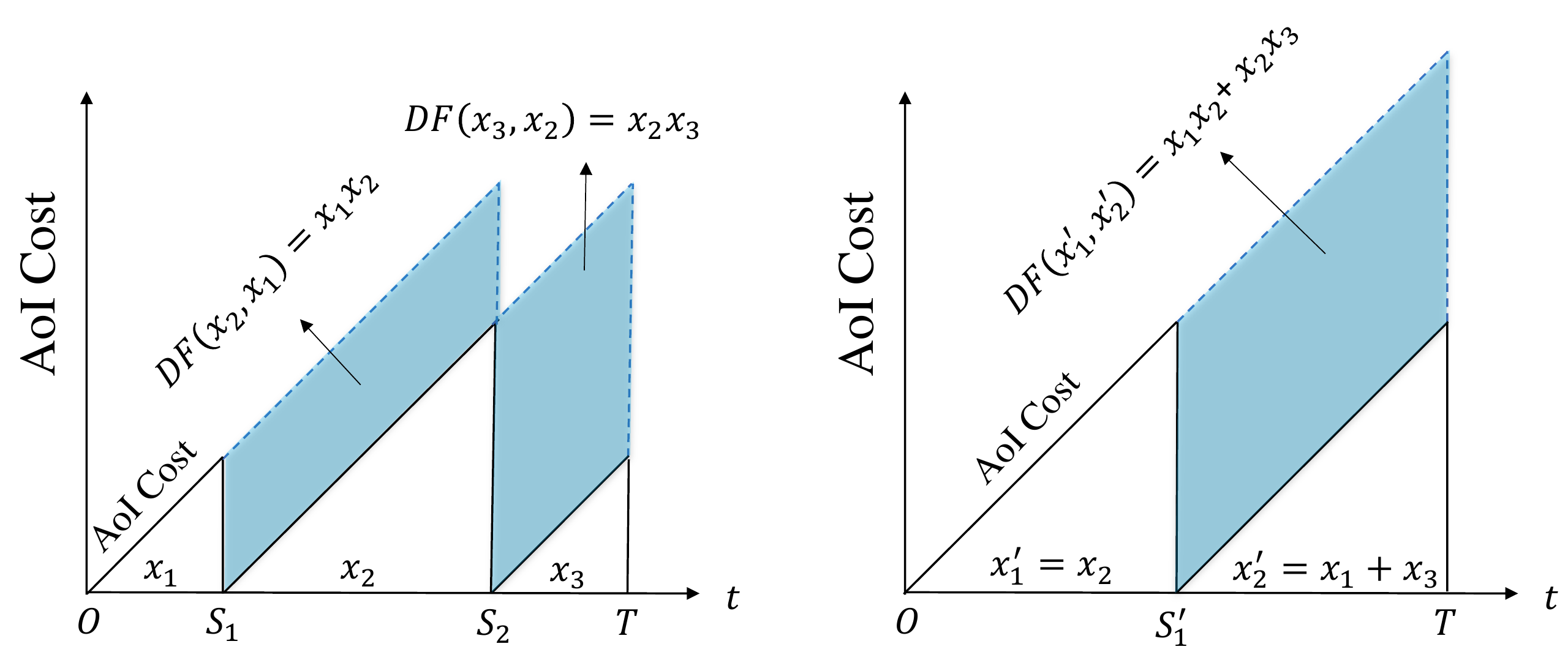}
		\vspace{-0.2cm}
		\caption{Illustrations of Example \ref{E1} with a linear cost function. Combining the first interval into the third interval maintains the payment.}
		\label{example1}
	\end{centering}
	\vspace{-0.3cm}
\end{figure}

\begin{example}\label{E1}
		Consider a linear AoI cost $f(\Delta_t)=\Delta_t$ and an arbitrary update policy $(K,\bs{x})$, as shown in Fig. \ref{example1}. For any time-dependent pricing scheme that induces only $K\geq 2$ updates. \rev{We will prove by induction that there exists a time-dependent pricing inducing $K-1$ updates and is more profitable.}
		
		\begin{itemize}
			\item \textbf{Base case:} When there are $K=2$ updates, as shown in Fig. \ref{example1}, the source's profit (the objective value in \eqref{Refor1}) is
			$x_1x_2+x_2x_3-C(2).$
			Consider another update policy $(1,x_1',x_2')$ where $x_1'=x_2$ and $x_2'= x_3+x_1$. The objective value  in \eqref{Refor1} becomes
			$x_2(x_1+x_3)-C(1).$
			Comparing these two values, 
			we see that $(1,x_1',x_2')$ is strictly more profitable than $(K,\bs{x})$.
			\item  \textbf{Induction step:} Let $K\geq n$ and suppose the statement that, for an arbitrary $K$-update pricing, there exists a more profitable $(K-1)$-update pricing
			is true for $K=n$. The objective value in \eqref{Refor1} is
			$\sum_{k=1}^Kx_kx_{k+1}-C(K).$
			Consider another update policy $(K'=K-1,\bs{x}')$ where $x_1'=x_2$, $x_2'= x_3+x_1$, and $x_k'=x_{k+1}$ for all other $k$. The objective value in \eqref{Refor1} becomes $
			(x_1+x_3)(x_2+x_4)+\sum_{k=4}^Kx_kx_{k+1}-C(K-1),$
			which is strictly larger than $P$. It is then readily verified that $(K',\bs{x}')$ is strictly more profitable than $(K,\bs{x})$. Based on induction, we can show that we can find a $(K'-1)$-update policy that is more profitable than the $(K',\bs{x}')$ policy. This eventually leads to the conclusion that a single update policy is the most profitable. 
		\end{itemize}
	\end{example}
	Based on the above technique,  we can show that the above argument works for any increasing convex AoI cost function. 
	\rev{We present the complete proof in Appendix \ref{ProofP2}.}

From Proposition \ref{P222}, it is readily verified that the optimal time-dependent pricing scheme is:
 	\begin{corollary}\label{P11}
		Under a convex AoI function $f(x)$, there exists an optimal time-dependent pricing scheme $\Pi^{*}_t$ such that\footnote{There actually exist multiple optimal pricing schemes; the only difference among all optimal pricing schemes are the prices for time instances other than $T/2$, which can be arbitrarily larger than $DF(T/2,T/2)$.}
		\begin{align}
		p^{*}({t})=DF\left(\frac{T}{2},\frac{T}{2}\right),~\forall t\in\mathcal{T},
		\end{align}
		where the equilibrium update takes place at $S_1^{\rm *, T}=T/2$.
	\end{corollary}
	We present the proof of Corollary \ref{P11} in Appendix \ref{ProofCoro1}.
	Corollary \ref{P11} suggests that there exists an optimal time-dependent pricing scheme that is in fact \textit{time-invariant}. That is, although our original intention is to exploit the time sensitivity/flexibility of the destination through the time-dependent pricing, it turns out not to be very effective. This motivates us to consider a quantity-based pricing scheme next.\footnote{The above analysis in Propositions \ref{P11} and \ref{P222} relies on the convex AoI cost function assumption. The analysis here for a general AoI cost function is difficult due to the resulted non-convexity of the problem in \eqref{Refor}. However, we will show that the optimal quantity-based and subscription-based pricing schemes are optimal among all pricing schemes under the general AoI cost functions.}
	
\subsection{Quantity-Based Pricing Scheme}

In this subsection, we focus on a quantity-based pricing scheme $\Pi_q=\{p_k\}_{k\in\mathbb{N}}$, i.e., the price depends on how many updates have been requested. Specifically, the price $p_k$ represents the price for the $k$-th update. The payment to the source is then given by $P(\mathcal{S},\Pi_q)=\sum_{k=1}^Kp_k.$

The source determines the quantity-based pricing scheme 
$\Pi_q$ in Stage I.
Based on $\Pi_q$,
the destination in Stage II chooses its update policy $(K,\bs{x})$.
We derive the \textit{(Stackelberg)} price-update equilibrium using the  bilevel optimization framework \cite{bilevel}. Specifically,
the bilevel optimization embeds the optimality condition of the destination's problem \eqref{destination-1} in Stage II into the  source's problem \eqref{source-1} in Stage I.
We first characterize the  conditions of the destination's update policy $(K^{\rm *}(\Pi_q),\bs{x}^{\rm *}(\Pi_q))$ that minimizes its overall cost in Stage II, based on which we
characterize the source's optimal pricing $\Pi_q^{*}$ in Stage I.\footnote{We use  $(K^{\rm *, Q},\bs{x}^{\rm *, Q})$ to denote the equilibrium update policy under the optimal quantity-based pricing, i.e., $(K^{\rm *, Q},\bs{x}^{\rm *, Q})=(K^{\rm *}(\Pi_q^{*}),\bs{x}^{\rm *}(\Pi_q^{*}))$.}

%
%

\subsubsection{Destination's Update Policy in Stage II} Given the quantity-based pricing scheme $\Pi_q$, the destination solves the following overall cost minimization problem:
\begin{subequations}\label{D}
	\begin{align}
	\min_{K\in\mathbb{N}\cup\{0\}, \bs{x}\in\mathbb{R}^{K+1}_{++}}~&\sum_{k=1}^{K+1} F(x_k)+\sum_{k=1}^Kp_k,\\
	{\rm s.t.}~~~~~~&\sum_{k=1}^{K+1}x_k=T.\label{D2}
	\end{align}
\end{subequations}
Note that the individual rationality constraint in \eqref{IRP} here is automatically satisfied, as the destination can always choose a no-update policy (i.e., $K=0$) leading to a cost of $F(T)$.
If we fix the value of $K$ in \eqref{D}, then problem \eqref{D} is convex with respect to $\bs{x}$. Such convexity allows to 
exploit the Karush–Kuhn–Tucker (KKT) conditions  in $\boldsymbol{x}$ to analyze the 
destination's optimal update policy in the following lemma:
\begin{lemma}\label{L4}
	Under any given  quantity-based pricing scheme $\Pi_q$ in Stage I, the destination's optimal update policy $(K^{\rm *}(\Pi_q),\bs{x}^{\rm *}(\Pi_q))$ satisfies
	\begin{align}
	x_k^{\rm *}(\Pi_q)=\frac{T}{K^{\rm *}(\Pi_q)+1},~~\forall k\in\mathcal{K}(K^{\rm *}(\Pi_q)+1). \label{L4eq}
	\end{align}
\end{lemma}
We present the proof of Lemma \ref{L4} in Appendix \ref{ProofL4}.
Intuitively, the KKT conditions of the problem in \eqref{D} equalize $f(x_k)$ for all $k$ and hence lead to the equal-spacing optimal update policy in \eqref{L4eq}.

\subsubsection{Source's Quantity-Based Pricing in Stage I}
Instead of solving ($K^{\rm *}(\Pi_q),\boldsymbol{x}^{\rm *}(\Pi_q)$) explicitly  in Stage II, we  apply the bilevel optimization  to solving the optimal quantity-based pricing $\Pi_q^{\rm *}$ in Stage I, which leads to the price-update equilibrium of our entire two-stage game \cite{bilevel}.
Substituting \eqref{L4eq} into the source's pricing in \eqref{source-1} yields the following bilevel problem: for all $k\in\mathcal{K}(K+1)$,
\begin{subequations}\label{Bilevel}
	\begin{align}
	{\rm \mathbf{Bilevel}:}~~\max_{\Pi_q, K, \bs{x}}&~~\sum_{k=1}^K p_k-C(K),\\
	{\rm s.t.}&~~K\in\arg\min_{K'\in\mathbb{N}\cup\{0\}}\Upsilon(K',\Pi_q)~,\nonumber\\
	&~~x_k=\frac{T}{K+1},~\label{C3}
	\end{align}
\end{subequations}
where $\Upsilon(K',\Pi_q)\triangleq(K'+1)F\left(\frac{T}{K'+1}\right)+\sum_{k=1}^{K'}p_k$ is the overall cost given the equalized interarrival time intervals.



We are now ready to present the optimal solution to
the bilevel optimization in \eqref{Bilevel}:
\begin{proposition}\label{T5}
	The equilibrium update count $K^{\rm *, Q}$ and the optimal  quantity-based pricing scheme ${\Pi}_q^{\rm *}$ satisfy
	\vspace{-0.3cm}
	\begin{align}
	\hspace{-2cm}	\sum_{k=1}^{K^{\rm *, Q}}p^{{\rm *}}_{k}&=F(T)-(K^{\rm *, Q}+1)F\left(\frac{T}{K^{\rm *, Q}+1}\right),\label{Ds}\\
	\sum_{k=1}^{K'}p^{{\rm *}}_{k}&\geq F(T)-(K'+1)F\!\left(\frac{T}{K'+1}\right),\forall K'\!\in\mathbb{N}\backslash\{K^{\rm *, Q}\}.\label{Da}
	\end{align}
\end{proposition}
We present the proof of Proposition \ref{T5} in Appendix \ref{ProofP3}.
Intuitively, 
the right-hand side of \eqref{Ds}
is the aggregate AoI cost difference between the no-update scheme and the optimal update policy.
Inequality \eqref{Da} together with \eqref{Ds} will  ensure that constraint \eqref{C3} holds. That is, if \eqref{Da} is not satisfied
or $\sum_{k=1}^{K^{\rm *, Q}}p_k^{*}>F(T)-(K^{\rm *, Q}+1)F\left(T/(K^{\rm *, Q}+1)\right)$,
then $K^{\rm *, Q}$
would violate constraint \eqref{C3}.
%
On the other hand, 
if $\sum_{k=1}^{K^{\rm *, Q}}p_k^{*}<F(T)-(K^{\rm *, Q}+1)F\left(T/(K^{\rm *, Q}+1)\right)$, then the source can always properly increase $p_1^{*}$ until \eqref{Ds} is satisfied. Such an increase does not violate constraint \eqref{C3} 
but improves the source's profit, contradicting with the optimality of $\Pi_q^{\rm *, Q}$. We will present an illustrative example of $\Pi_q^*$ in Section \ref{summary-P}.
%
%

Substituting the pricing structure in \eqref{Ds} into \eqref{Bilevel}, we can obtain  $K^{\rm *, Q}$ through solving the following problem:
\begin{align}
\max_{K\in\mathbb{N}\cup\{0\}}~-(K+1)F\left(\frac{T}{K+1}\right)-C(K).\label{PBK}
\end{align}

To solve problem \eqref{PBK}, we first relax the constraint $K\in\mathbb{N}\cup\{0\}$ into $K\in\mathbb{R}_+$,  and then \rev{recover the integer solution by rounding}.
We start with relaxing the integer constraint $K\in \mathbb{N}\cup\{0\}$ constraint in \eqref{PBK} into $K\geq 0$, which leads to a convex problem.\footnote{To see the convexity of 
$(K+1)F\left(T/(K+1)\right)$, note that $(K+1)F\left(T/(K+1)\right)$ is the \emph{perspective} function of $F(T)$. The perspective function of $F(T)$ is convex since $F(T)$ is convex.
}
We take the derivative of objective in \eqref{PBK} to
define  a threshold update count $\hat{K}$ satisfying
\begin{subequations}\label{hatK}
	\begin{align}
	f\left(\frac{T}{\hat{K}+1}\right)\frac{T}{\hat{K}+1}-F\left(\frac{T}{\hat{K}+1}\right)&\geq C'(\hat{K}),\\
	f\left(\frac{T}{\hat{K}+2}\right)\frac{T}{\hat{K}+2}-F\left(\frac{T}{\hat{K}+2}\right)&< C'(\hat{K}+1).
	\end{align}
\end{subequations}
\rev{Note that Assumption \ref{A2} leads to the existence of a unique $\hat{K}$ satisfying \eqref{hatK} and the convexity of the objective in \eqref{PBK} ensures that values other than these two candidates ($\hat{K}$ and $(\hat{K}+1)$) are not optimal to the problem  in \eqref{PBK}. Therefore, the threshold counts $\hat{K}$ and $(\hat{K}+1)$ serve as candidates for the optimal update count to  the problem  in \eqref{PBK} as shown next.}
\begin{proposition}\label{P3}
	The optimal  update count $K^{\rm *, Q}$ to problem in  \eqref{Bilevel} satisfies
	\begin{align}\label{eq28}
	K^{\rm *, Q}=	\arg\min_{K\in\{\hat{K},\hat{K}+1\}}(K+1)F\left(\frac{T}{K+1}\right)+C(K).
	\end{align} 
\end{proposition}
We present the proof of Proposition \ref{P3} in Appendix \ref{ProofP4}.
After obtaining $K^{\rm *, Q}$, we can construct an equilibrium pricing scheme based on Proposition \ref{T5}. An example optimal quantity-based pricing is
	\begin{align}\hspace{-0.5cm}\label{optimalpricing}
	&p^{\star}_k=\nonumber\\
	&\begin{cases}
	F(T)-2F(\frac{T}{2})+\epsilon~~~~~~~~~~~~~~~~~~~~~~~~~~~~~~~{\rm if}~k=1,\\
	F(T)-(k+1)F(\frac{T}{k+1})-\sum_{j=1}^{k-1}p^{\star}_j+\epsilon,~~~~~{\rm if}~1< k< K^{\rm *, Q},\\
	F(T)-(K^{\rm *, Q}+1)F(\frac{T}{K^{\rm *, Q}+1})-\sum_{j=1}^{K^{\rm *, Q}-1}p^{\star}_j,~{\rm if}~k\geq K^{\rm *, Q}.
	\end{cases}
	\end{align}
	where $\epsilon>0$ is infinitesimal to ensure \eqref{Da}. We present an illustrative example of \eqref{optimalpricing} in \cite{technical}.

We next show that the optimal quantity-based pricing scheme is in fact profit-maximizing among all possible pricing schemes. To see this,
note that \eqref{eq28} is socially optimal as it is equivalent to the SCM-F Problem in \eqref{SCM-F}. From Lemma \ref{LLL1}, the following is readily verified:
\vspace{-0.25cm}
\begin{theorem}[Surplus Extraction]\label{P2}
	The optimal quantity-based pricing $\Pi_q^{*}$ is surplus extracting, i.e., it
	achieves the maximum source profit
	among all possible pricing schemes.
\end{theorem}
\vspace{-0.25cm}
We present the proof of Theorem \ref{P2} in Appendix \ref{ProofT1}.
Theorem \ref{P2} implies that the quantity-based pricing scheme is already one of the optimal pricing schemes. 
Hence, even without exploiting the time flexibility explicitly, it is still possible to obtain the optimal pricing structure, which again implies that utilizing time flexibility may be not necessary under the finite-horizon model.

%



\subsection{Subscription-Based Pricing}

In this subsection, we consider a subscription-based pricing $\Pi_s=\{\pi,p_u \}\in \mathbb{R}_+^2$, where $\pi$ is a \textit{one-time subscription price} and $p_u$ corresponds to a (fixed-rate) \textit{usage price} for each update. That is, for an update policy with $K$ updates, the payment is $P(\mathcal{S},\Pi_s)=\pi+K\cdot p_u$. Compared to the quantity-based pricing and the time-dependent pricing, such a pricing scheme enjoys a low implementation complexity as it is characterized by two variables only.

\rev{Recall that the surplus-extracting pricing (in Definition \ref{Def7}) leads to a socially optimal update policy.
Hence, the key idea of constructing the subscription-based pricing is to set $p_u$ to induce socially optimal update policy and then charges the maximal $\pi$ that satisfies the individual rationality constraint in \eqref{IRP}.} We now have the following result:
\begin{proposition}\label{P22}
	{Let $(K^{o},x^o)$ be the socially optimal update policy solving the SCM-F Problem in \eqref{SCM-F}).}
	The following  subscription-based pricing $\Pi_s^*=\{\pi^*,p_u^* \}$ is surplus-extracting:
	\begin{subequations}
	\begin{align}\label{subscrip}
	\pi^{\rm *}&=F(T)-(K^{o}+1)F\left(\frac{T}{K^{o}+1}\right)-c(x^o)K^{o}\\
	p_u^{\rm *}&=c(x^o).
	\end{align} 
	\end{subequations}
\end{proposition}
Before discussing the reason why the pricing scheme in \eqref{subscrip} can achieve the maximal profit, we first note that the optimal subscription pricing is a special case of the optimal quantity-based pricing.
We construct an equivalent quantity-based pricing (yielding the same source's profit) satisfying Proposition \ref{T5}, $\hat{\Pi}_q=\{\hat{p}_k\}_{k\in\mathbb{N}}$ via
\begin{align}\label{eq7}
\hat{p}_k=
\begin{cases}
p_u^{\rm *}+\pi^{\rm *},~~~~~~&{\rm if}~~k=1,\\
\pi^{\rm *},~~~~~~~~~~&{\rm otherwise}.
\end{cases}
\end{align}
Substituting \eqref{eq7} into Proposition \ref{T5}, we see that $\hat{\Pi}_q$ is the optimal quantity-based pricing, which is surplus-extracting by Theorem \ref{P2}.

Although the optimal subscription-based pricing scheme corresponds to a special case of the optimal quantity-based pricing scheme under the finite-horizon model, it is not the case in the infinite-horizon model, as we will analyze in Section \ref{SBP-U}. 
\subsection{Summary}\label{summary-P}

\begin{figure}[t]
	\centering
	\subfigure[Optimal time-dependent pricing]{\includegraphics[scale=.21]{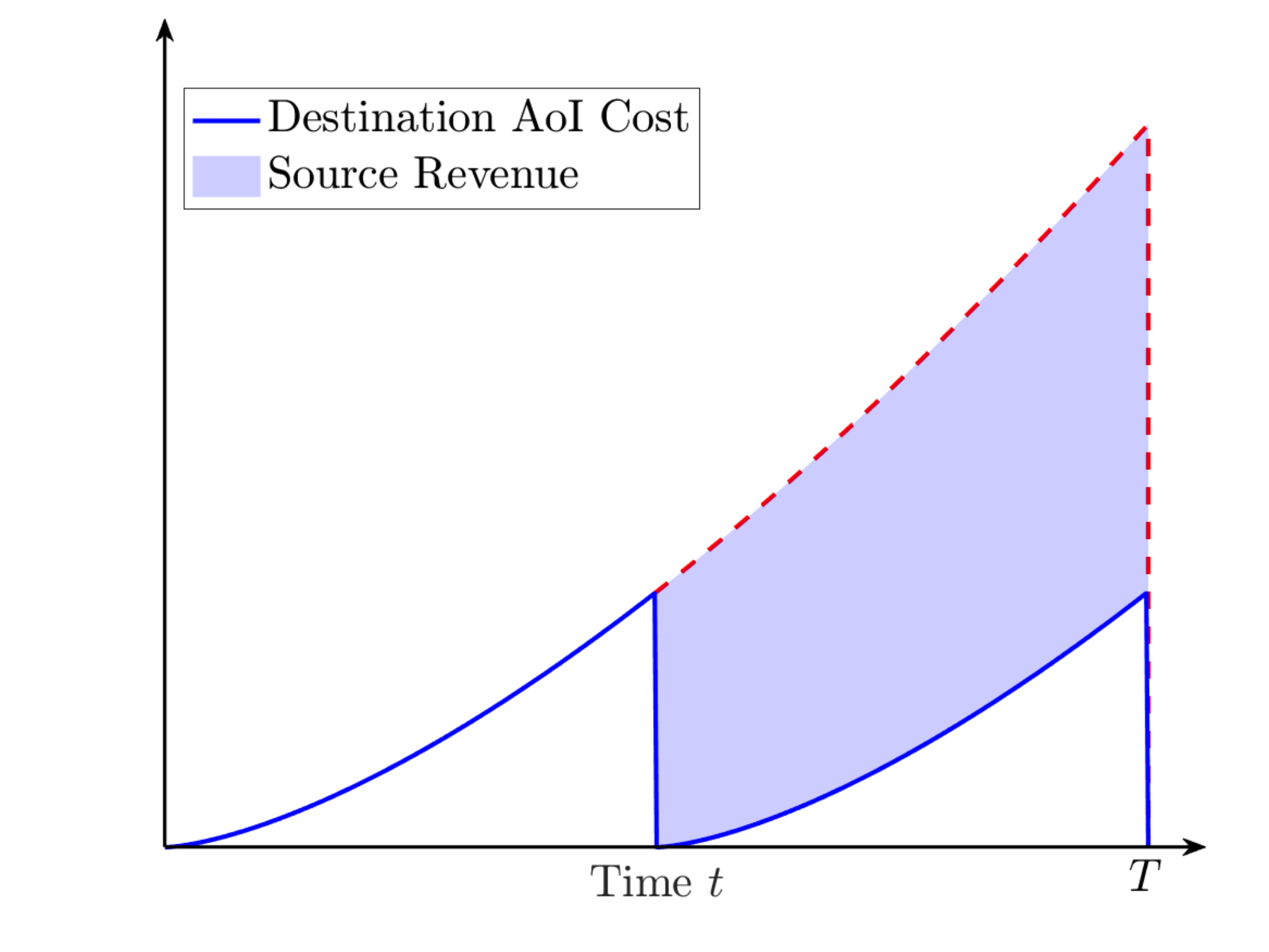}}
		\subfigure[Optimal quantity-based pricing and subscription-based pricing]{\includegraphics[scale=.21]{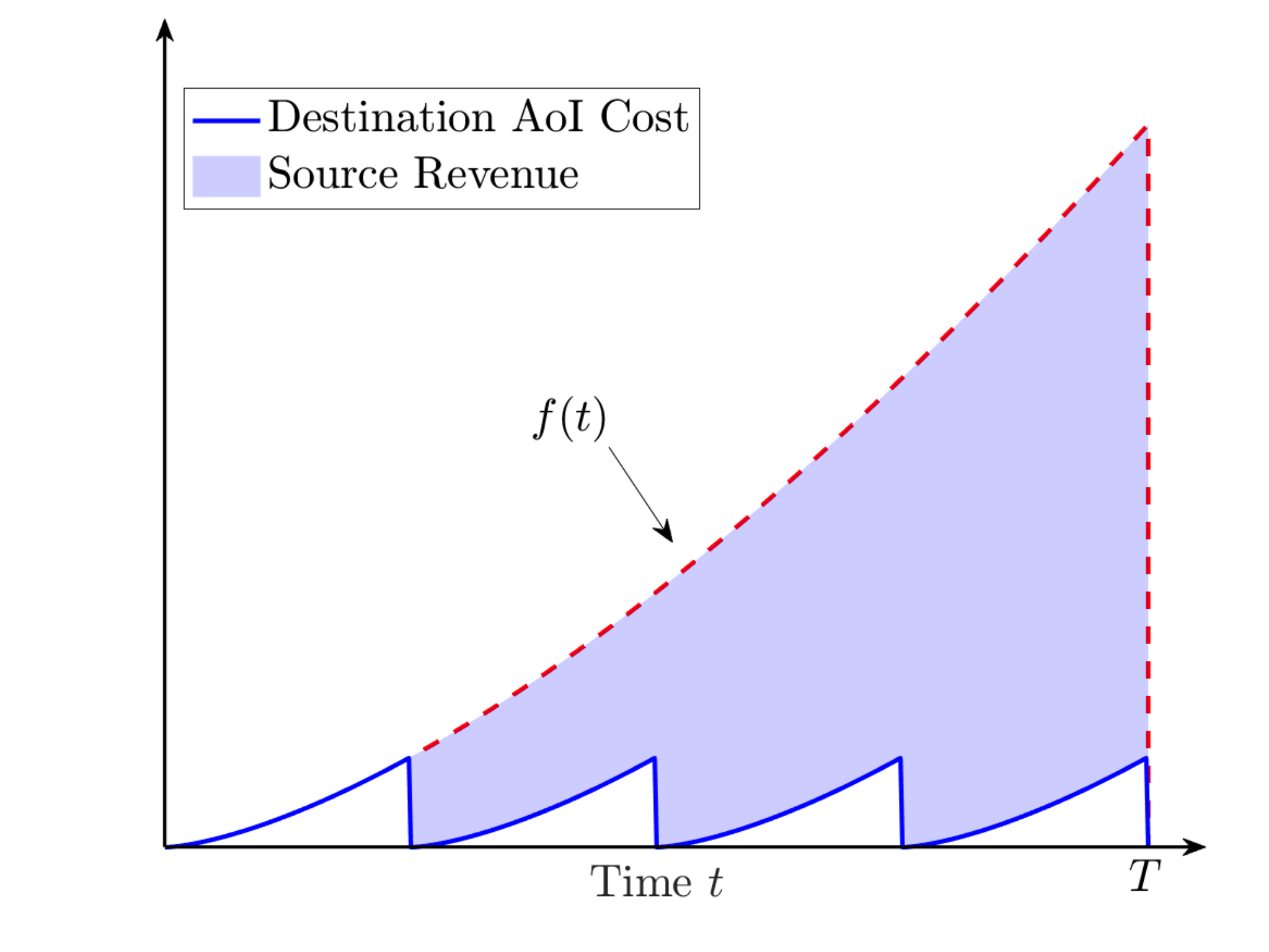}}
		\vspace{-0.3cm}
		\caption{Performance comparison in terms of the AoI cost and the revenue under a convex AoI cost.}
		\label{Sum1}
\end{figure}

To summarize our key results in this section, we graphically compare the AoI costs and the revenues under three studied pricing schemes in Fig. \ref{Sum1} under a convex AoI cost. As in Fig. \ref{Sum1}(a), the optimal time-dependent pricing scheme generates a revenue for the source equal to the differential aggregate AoI cost (Lemma \ref{L1}), and induces a unique update at $T/2$ (Proposition \ref{P222}). We present the results regarding the optimal quantity-based pricing and the subscription-based pricing in Fig. \ref{Sum1}(b), since the latter corresponds to a special case of the former, as shown in \eqref{eq7}.
The generated revenue equals the difference of the aggregate AoI costs under a no-update policy and the social optimum update policy. Finally, both the optimal quantity-based pricing and the subscription-based pricing are surplus-extracting (Theorem \ref{P2} and Proposition \ref{P22}) and thus maximize the source's profit among all possible pricing schemes (Lemma \ref{LLL1}).

\section{{Infinite-Horizon Model}}\label{Unpredict}

We now analyze the 
infinite-horizon model, in which valuations and costs are discounted over time. 
Specifically, the source's and the destination's decisions account for \textit{time discounting}: the source and the destination discount payments and costs as they approach a temporal horizon into the future \cite{discount}. 
 This renders the analysis more challenging, since
the destination and the sources' problems become non-convex continuous-time dynamic programs.

We aim at designing three pricing schemes and compare their performances, and we \rev{will show that they will behave differently compared with the finite-horizon model.} {To distinguish between notations in both models, we
use superscript $\star$ to indicate the equilibrium notations under the infinite-horizon model.}
\subsection{Problem Formulation}
The analysis in the infinite-horizon model is significantly different from that in the finite-horizon model, mainly due to the time discounting.
We denote by $\delta$ the \textit{discount coefficient}, which corresponds to the level that the payment and the cost are
discounted after each unit of time. We then introduce the following notations:
\begin{definition}[Discounted Notations]
The \textit{discounted payment} $P_\delta(\mathcal{S},\Pi)$, the source's \textit{discounted operational cost} $C_\delta(\mathcal{S})$, and the destination's \textit{discounted aggregate and cumulative AoI costs} $\Gamma_\delta(\mathcal{S})$ and $F_{\delta}(x)$ are
\begin{align}
    P_\delta(\mathcal{S},\Pi)&\triangleq \pi+\sum_{k=1}^{\infty}\delta^{S_k}p_k(S_k),~~ C_\delta(\mathcal{S})\triangleq \sum_{k=1}^{K}\delta^{S_k}c(\bar{x}),\\
        \Gamma_\delta(\mathcal{S})&\triangleq\int_{0}^\infty \delta^t f(\Delta_t(\mathcal{S}))dt,~{\rm and}~ F_{\delta}(x)\triangleq\int_0^x\delta^tf(t)dt,
\end{align}
where the average interarrival time $\bar{x}$ is now given by $\bar{x}=\lim_{K\rightarrow\infty}(\sum_{k=1}^{K}S_k-S_{k-1})/K.$
\end{definition}
%




The \textit{individual rationality} constraint in pricing scheme design is then given by:
\vspace{-0.2cm}
\begin{align}
    \Gamma_\delta(\mathcal{S}^{\star}(\Pi))+P_\delta(\mathcal{S}^{\star}(\Pi),\Pi)\leq \Gamma_\delta(\infty), \label{IRU}
\end{align}
where $\mathcal{S}^{\star}(\Pi)$ is the destination's optimal update policy to be defined in the following.
\vspace{-0.25cm}
\begin{game}[Source-Destination Interaction Game with Time Discounting] The source and the destination interact in the following two stages:
\begin{itemize}
    \item  In Stage I, the source determines the pricing scheme function $\Pi$ at the beginning of the period, in order to maximize its \textbf{discounted profit} as follows:
\vspace{-0.2cm}
\begin{subequations}\label{source-2}
	\begin{align}
	&{\rm \mathbf{Source-I:}}~\nonumber\\
	&\max_{\Pi}~P_\delta(\mathcal{S}^{\star}(\Pi),\Pi)-C_\delta(\mathcal{S}^{\star}(\Pi)),\\
	&~{\rm s.t.}~~\Pi\in\{\Pi: \eqref{IRU}, \pi, p_k(t)\geq0,~\forall t\in\mathcal{T}, k\in\mathbb{N}\}.
	\end{align}
\end{subequations}
\item In Stage II, the destination decides its update policy 	to minimize its discounted aggregate AoI cost plus discounted payment:
\begin{align}\label{destination-2}
&{\rm \mathbf{Destination-I:}}\nonumber\\ &~\mathcal{S}^{\star}(\Pi)=\arg\min_{\mathcal{S}\in\Phi}~\Gamma_\delta(\mathcal{S})+P_\delta(\mathcal{S},\Pi).
\end{align}
\end{itemize}
\end{game}

\subsection{Social Cost Minimization and Surplus Extraction}
In this subsection, we present the social optimum update policy and the surplus-extracting profit as a upper bound for the source's achievable profit.
We start with defining the \textit{Discounted Social Cost Minimization (SCM-I) problem} as follows:
\begin{subequations}\label{D-SCM}
	\begin{align}
	{\rm \mathbf{SCM-I:}}\nonumber\\
	~\min_{\mathcal{S}\in\Phi}\quad &
	F_{\delta}(S_1)+\lim_{K\rightarrow \infty} \sum_{k=1}^{K}\delta^{S_{k}} \left[F_{\delta}(S_{k+1}-S_{k})+c\left(\bar{x}\right)\right].
	\end{align}
\end{subequations}
The SCM-U Problem is a \textit{continuous-time dynamic programming} problem, which can be tackled by breaking into a sequence of decision steps over time. To do so, we 
let $V_c$ denote the minimal social cost (the minimal objective value of the SCM-U Problem) and introduce the following result towards solving the SCM-U Problem:
\begin{lemma}\label{L5} The minimal social cost $V_c$ satisfies
	\begin{align}
	&V_c=\nonumber\\
	&\min_{S_k}\left\{F_\delta(S_k-S_{k-1})+\delta^{S_k-S_{k-1}} [c(S_k-S_{k-1})+V_c]\right\},\nonumber\\
	&~{\rm s.t.}~S_k\geq S_{k-1}, \forall k\in\mathbb{N}.\label{MSC}
	\end{align}
\end{lemma}

We present the proof of Lemma \ref{L5} in Appendix \ref{ProofL5}. 
{ Lemma \ref{L5} implies that the optimization problem to be solved at $t=S_k$ is similar to that at $t=0$, which implies that the optimal solution $\mathcal{S}^o$ is in fact 
stationary and hence is \textit{equal-spacing}.} {Taking the derivative of \eqref{MSC} yields the following result:}
\begin{proposition}\label{P5}
	The social cost minimizing policy $\mathcal{S}^o$ satisfies $
	    S_k^o=kx^o,~\forall k\in\mathbb{N},$
	where $x^o$ is the \textit{socially optimal interarrival time} satisfying 
\begin{align}
&\int_0^{x^o}(1-\delta^t)f'(t)dt\nonumber\\
=&\ln(\delta^{-1})\left[c(x^o)-\int_{0}^{x^o}\left(\ln(\delta)\delta^t c'(t)+(1-\delta^t) c''(t)\right)dt
    \right].\label{P5Eq34}
\end{align} 
\end{proposition}
We present the proof of Proposition \ref{P5} in Appendix \ref{ProofP5}.
The left-hand side of \eqref{P5Eq34} is increasing in $x^o$ and the right-hand side is decreasing in $x^o$, which implies that $x^o$ is uniquely defined and can be efficiently obtained by the bisection method.
Finally, we derive the upper bound for source's profit analog to the finite-horizon model:
\begin{definition}[Surplus Extraction]\label{Def3}
A pricing scheme $\Pi$ is surplus-extracting if it satisfies
\begin{align}
P_\delta(\mathcal{S}^{\star}(\Pi),\Pi)=F_\delta(\infty)-\Gamma_\delta(\mathcal{S}^{\star}(\Pi))~{\rm and}~ \mathcal{S}^{\star}(\Pi)=\mathcal{S}^o,\label{SE}
\end{align}
and $\mathcal{S}^{o}$ is socially optimal, i.e., solves \eqref{D-SCM}.
\end{definition}
To ensure that $F_\delta(\infty)$ is finite, we adopt the following assumption throughout this paper:
\begin{assumption}\label{Assum2}
There exists parameters $A, \zeta, \gamma$ satisfying $A<\infty$, $\zeta\delta\leq \gamma < 1$, and
$f(t)\leq A \zeta^t,~\forall t\geq 0.
$
\end{assumption}
Assumption \ref{Assum2} prevents $\delta^tf(t)$ from diverging to $\infty$. It is satisfied by many classes of AoI functions including concave AoI functions and polynomial AoI functions (as in \cite{AoI7}).
Assumption \ref{Assum2} further ensures that $F_\delta(\infty)$ is finite, since $F_\delta(\infty)=\int_{0}^\infty \delta^tf(t) dt\leq A \int_{0}^\infty (\delta\zeta)^t dt=A/\ln((\delta\zeta)^{-1})$. Based on the proof technique 
similar to that of Lemma \ref{LLL1}, we have
\begin{lemma}\label{LLL2}
When Assumption \ref{Assum2} is satisfied, a surplus-extracting pricing scheme is the optimal pricing among all possible pricing schemes under the infinite-horizon model.
\end{lemma}
We present the proof of Lemma \ref{LLL2} in Appendix \ref{ProofL555}.

\subsection{Time-Dependent Pricing Scheme}

In this subsection, we study the time-dependent pricing scheme $\Pi_t=\{p(t)\}_{t\geq 0}$ under the infinite-horizon model. Based on our analysis of the time-dependent pricing under the finite-horizon model, we derive an equilibrium condition.
Although the time-dependent pricing scheme is also not surplus-extracting and the corresponding optimization is difficult to solve, we present a suboptimal solution and show its asymptotic surplus-extraction.

\subsubsection{Equilibrium Condition}
Recall that the time-dependent pricing design under the finite-horizon model is based on the differential AoI cost. We next introduce the similar result for the infinite-horizon, analog to Lemma \ref{L1}. 
\begin{lemma}\label{L6}
	Any equilibrium price-update pair $(\Pi_{t}^{\rm \star},\mathcal{S}^{\rm 
	\star, T}\triangleq \mathcal{S}^{\rm 
	\star}(\Pi_{t}^{\rm \star}))$ should satisfy, for all $k\in\mathbb{N}$,
	\begin{align}
	p^{\star}\left(S_k^{\rm \star, T}\right)=& F_\delta(S_{k+1}^{\rm \star, T}-S_{k-1}^{\rm \star, T})-F_\delta(S_{k}^{\rm \star, T}-S_{k-1}^{\rm \star, T})\nonumber\\
	&-\delta^{S_{k}^{\rm \star, T}-S_{k-1}^{\rm \star, T}}F_\delta(S_{k+1}^{\rm \star, T}-S_{k}^{\rm \star, T}).\label{price2}
	\end{align}
\end{lemma}
The intuition is similar to the optimal time-dependent pricing scheme discussed previously, i.e.,
the right hand side of \eqref{price2} equals  the destination's  maximal willingness to pay. 
For all time instances other than ${S}^{\rm \star, T}_k$ for all $k$, the source can impose infinitely large prices to
ensure that  the destination does not update at any of these time instances.
Lemma \ref{L6} enables us to reformulate time-dependent pricing scheme  into the following dynamic programming problem:
\begin{align}
\max_{\mathcal{S}\in\Phi}~&\sum_{k=1}^\infty\delta^{S_{k-1}}[F_\delta(S_{k+1}-S_{k-1})\! -\! F_\delta(S_{k}-S_{k-1})]\nonumber\\
&-\delta^{S_{k-1}}F_\delta(S_{k+1}-S_k)-\delta^{S_k}c(\bar{x})\label{TDP-U}.
\end{align}
Solving problem \eqref{TDP-U} requires us to analytically derive a value function, which is challenging. This motivates us to consider a suboptimal time-dependent pricing scheme next.
\subsubsection{Suboptimal Time-Dependent Pricing and Algorithm}
\rev{Motivated by the fact that the surplus-extracting pricing scheme in Definition \ref{Def3} is equal-spacing,}
we will next search for a (suboptimal) equal-spacing  time-dependent pricing scheme
by solving the following problem:
\begin{align}
 \max_{x\geq 0}	\frac{F_{\delta}(2x)-(1+\delta^x)F_\delta(x)-\delta^xc(x)}{1-\delta^x}. \label{suboptimal}
\end{align}
In \eqref{suboptimal}, the scalar variable $x$ denotes the interarrival time between each adjacent updates and we derive the discounted profit based on Lemma \ref{L6}.
\rev{The problem in \eqref{suboptimal} is much more tractable than \eqref{TDP-U} since it only requires solving an one-dimensional optimization problem.}

%
%
   
%


To solve the above problem in \eqref{suboptimal}, we will adopt the fractional programming technique in \cite{FP} by 
introducing the following problem:
\begin{align}
    \max_{x\geq 0}~\mathcal{L}(x,Q)\triangleq &F_{\delta}(2x)-(1+\delta^x)F_\delta(x)-\delta^xc(x) \nonumber\\
    &-Q\cdot(1-\delta^x).\label{FP}
\end{align}
Let $Q^\star$ be the maximal objective value of \eqref{suboptimal}.
From \cite{FP}, $Q^\star$ and the optimal solution $x_t^\star$ to the problem in \eqref{suboptimal} should satisfy
\begin{align}
   \max_{x}\mathcal{L}(x,Q^\star)=0, \quad{\rm and}\quad x^\star_{t}=\arg\max_{x\geq 0}  \mathcal{L}(x,Q^\star). \label{lambda*}
\end{align}
 It is readily verified that $\max_{x}\mathcal{L}(x,Q)$ is decreasing in $Q$, 
 which implies that we can adopt the bisection search for $Q^\star$ once we can solve the problem in \eqref{FP} for every $Q>0$. Therefore, to obtain $x_t^\star$, we first fix $Q$ and solve the problem in \eqref{FP}, and then search for $Q$ satisfying \eqref{lambda*}.

Although the problem in \eqref{FP} is non-convex, a brute-force  one-dimensional search with the time complexity of $\mathcal{O}(M)$ in fact leads to the close-to-optimal solution to problem \eqref{FP}, to be shown next.
Algorithm \ref{Algo1} summarizes the above procedure. Lines \ref{l4} and \ref{l7}-\ref{l10} perform the bisection search for $Q^\star$ and Line \ref{LineFP} performs the brute-force search for the optimal solution to \eqref{FP}.

\begin{algorithm}[tb]
	\SetAlgoLined
	\caption{Dinkelbach Method to solve \eqref{suboptimal}}\label{Algo1}
	\footnotesize{Initialize the number of samples $M$, the iteration index $n$, $Q_{\rm L}$, $Q_{\rm H}$, and a tolerance parameter $\epsilon>0$\;
		\While{ $| Q_{\rm H}-Q_{\rm L}|\geq \epsilon$ }{
		   Set $n= n+1$ and $Q[n]=\frac{Q_{\rm H}+Q_{\rm L}}{2}$\; \label{l4}
		    Generate a sequence of $\mathcal{X}_T\triangleq\{\frac{k\tilde{x}(Q[n])}{M}\}_{k\in\{1,2,...,M\}}$\;
            Find $x[n]$ such that $x[n]\in\arg \max_{x\in\mathcal{X}_T} \mathcal{L}(x,Q[n])$\label{LineFP}\;
				\eIf{$\mathcal{L}(x[n],Q[n])>0$\label{l7}}{
				 Set $Q_{\rm L}=Q[n]$\;
				}{ Set $Q_{\rm H}=Q[n]$\;\label{l10}}
				
		}
		}
\end{algorithm}

To show the optimality of Algorithm \ref{Algo1} in terms of solving \eqref{FP}, we define $\hat{x}\triangleq \arg_x\{\int_{0}^x \delta^t[f(x+t)-f(t)]dt=c(x)   \}$\footnote{Note that $\hat{x}$ always exists and is uniquely defined since $\int_{0}^x \delta^t[f(x+t)-f(t)]dt$ is continuous and increasing in $x$, $c(x)$ is continuous and non-increasing in $x$, and $\int_{0}^x \delta^t[f(x+t)-f(t)]dt=0$ when $x=0$ and $\lim_{x\rightarrow \infty}\int_{0}^x \delta^t[f(x+t)-f(t)]dt\rightarrow \infty$. } as the interarrival time yielding a zero objective value of \eqref{suboptimal}. We define 
\begin{align}
    \tilde{x}(Q)\triangleq \hat{x}+\log_\delta\left(\frac{Q\ln((\delta\zeta)^{-1})}{1+Q\ln((\delta\zeta)^{-1})}\right)+\log_\delta(\delta\cdot\max(1,\zeta)).\label{tildex}
\end{align}
We are ready to present the following result showing that the objective value loss of $\eqref{FP}$ diminishes in $M$ (the number of samples in Algorithm \ref{Algo1}):
\begin{proposition}\label{P8}
Algorithm \ref{Algo1} in Line \ref{LineFP} yields an solution $x[n]$ to the problem in \eqref{FP} such that
 $$\max_{x\geq 0} \mathcal{L}(x,Q[n])-\mathcal{L}(x[n],Q[n])\leq \frac{L_{\mathcal{L}}(Q[n])\tilde{x}(Q[n])}{2M},$$
 where $L_\mathcal{L}(Q[n])\triangleq 4\max_{t\geq 0}[\delta^tf(t)]+L_c+Q[n] \ln(\delta^{-1})$ is the Lipschitz constant of $\mathcal{L}(x,Q[n])$.
\end{proposition}
 We present the proof of Proposition \ref{P8} in Appendix \ref{ProofP7}.
Hence, Algorithm \ref{Algo1} generates an (approximately) optimal solution $x[n]$ to \eqref{FP} with an $\mathcal{O}(1/M)$ objective value loss. The proof of Proposition \ref{P8} involves showing the existence of the optimal solution to \eqref{FP} in $[0,\tilde{x}(Q)]$ and the Lipschitz continuity of $\mathcal{L}(x,Q)$ in $x$.

Finally, from Lemma \ref{L6}, the equal-spacing time-dependent pricing scheme $\tilde{\Pi}_t=\{\tilde{p}(t)\}_{t\geq 0}$ based on the optimal solution to the problem in \eqref{FP} is
	\begin{align}\label{STDP}
	\tilde{p}\left(t\right)=\begin{cases}
	F_\delta(2x^\star_{t})-(1+\delta^{x^\star_{t}})F_\delta(x^\star_{t}),&~{\rm if}~t=kx^\star,~k\in\mathbb{N},\\
	+\infty,&~{\rm otherwise}.
	\end{cases}
	\end{align}

\subsubsection{Asymptotic Surplus-Extraction}

We next study how profitable such a suboptimal time-dependent pricing can be, through the following proposition:
\begin{proposition}\label{Asy}
    The suboptimal time-dependent pricing in \eqref{STDP} is asymptotically surplus-extracting as $\delta\rightarrow 0$.
\end{proposition}	
We present the proof of Proposition \ref{Asy} in Appendix \ref{ProofP8}.
Proposition \ref{Asy} shows that the suboptimal time-dependent pricing scheme is in fact close-to-optimal among all pricing schemes when $\delta$ is small enough. Hence, it implies that exploiting the time dimension is profitable when the source and the destination are ``impatient'', even though the time-dependent pricing scheme is not effective in the finite-horizon model as discounting is not considered there.

\subsection{Quantity-Based Pricing Scheme}

In this subsection, we consider the quantity-based pricing scheme $\Pi_q$, \rev{i.e., instead of differentiating the prices across time, the price for each update changes
as the destination requests more.} We will study whether the optimal quantity-based pricing is still surplus-extracting as it is in the finite-horizon model. We commence with the destination's update policy analysis.

\subsubsection{Destination's Update Policy in Stage II}




We first define $\tilde{\Pi}_{q,j}\triangleq \{\tilde{p}_{k,j}\}_{k\in\mathbb{N}}$ such that $\tilde{p}_{k,j}=p_{k+j}$ for all $k$ and all $j$. {We further define $f_\delta(x)$ as the discounted AoI, given by $f_\delta(x)\triangleq\delta^x f(x)$.}
To characterize the destination's update policy $\mathcal{S}^{\star}(\Pi_q)$ under an arbitrary quantity-based pricing scheme $\Pi_q$, we consider the following lemma: 
\begin{lemma}\label{L66}
There exists \textit{a value function $V_q(\Pi_q)$}, representing the minimal destination's overall cost, that has the following recurrent form:
\begin{align}\label{AValue}
& V_q(\tilde{\Pi}_{q,k-1})\triangleq\nonumber\\
&~~~~\min_{S_k}~[F_\delta(S_k-S_{k-1}) +\delta^{S_k-S_{k-1}}(p_k+V_q(\tilde{\Pi}_{q,k}))],\nonumber\\
&~~~~{\rm s.t.}~S_k\geq S_{k-1}, \forall k\in\mathbb{N}.
\end{align}
Under any quantity-based pricing scheme $\Pi_q$, the destination's optimal update policy $\mathcal{S}^{\star}(\Pi_q)$ satisfies that
\begin{align}
&f_\delta(S_k^{\rm 
\star}(\Pi_q)-S_{k-1}^{\rm \star} (\Pi_q)) \nonumber\\
= &\ln(\delta^{-1})\delta^{S_k^{\rm \star} (\Pi_q)-S_{k-1}^{\rm \star}(\Pi_q)}(p_k+V_q(\tilde{\Pi}_{q,k})),~\forall k\in\mathbb{N}.\label{asda}
\end{align}
\end{lemma}

We present the proof of Lemma \ref{L66} in Appendix \ref{ProofL7}.
{Intuitively, for each update $k$, the destination selects the interarrival time to balance the discounted cumulative AoI cost $F_\delta(S_k-S_{k-1})$ and the delay of the future overall cost $(p_k+V_q(\tilde{\Pi}_{q,k}))$.} Note that it is difficult to obtain the exact form of the destination's value function in \eqref{AValue}. However, we will show that the optimality condition in Lemma \ref{L66} is sufficient for designing the optimal quantity-based pricing, as we will show next.
		
\subsubsection{Source's Pricing Design in Stage I} 

{Substituting the destination's update policy in Lemma \ref{L66} into the source's pricing problem in \eqref{source-2}, we can transform \eqref{source-2} into the following form:}
	\begin{align}
	&\max_{\mathcal{S}\in\Phi}~ \frac{1}{\ln(\delta^{-1})} f_\delta(S_1)-\lim_{K\rightarrow \infty} \sum_{k=1}^{K}\delta^{S_{k}} \left[F_{\delta}(S_{k+1}-S_{k})+c(\bar{x})\right]\label{PP},
	\end{align}
which leads to the destination's equilibrium update policy $\mathcal{S}^{\star,Q}\triangleq\mathcal{S}^{\star}(\Pi_q^{\star})$. Solving the problem in \eqref{PP} leads to the optimal quantity-based pricing 
$\Pi_q^{\star}=\{p_k^{\star}\}_{k\in\mathbb{N}}$ based on Lemma \ref{L66}.

In the following, we analytically solve the problem in \eqref{PP}.
We observe that the discounted social cost (defined in \eqref{D-SCM}) appears in the source's objective in \eqref{PP}. Based on such an observation, we can derive the following result towards solving the problem in \eqref{PP}:
\begin{lemma}\label{L8}
The update policy $\mathcal{S}^{\star, Q}$ that is optimal to the problem in \eqref{PP} should satisfy
\begin{align}
&S_k^{\star, Q}=\nonumber\\
&\begin{cases}\arg \max_{S_k\geq 0}\!\!\left[\frac{f_{\delta}(S_k)}{\ln(\delta^{-1})}\!-\!\delta^{S_k} (c(x^o)+V_c)\right],&{\rm if}~k=1,\\
S_{k-1}^o+S_1^{\star, Q},&{\rm otherwise},
\end{cases}
\label{Bell}
\end{align}
{where $\mathcal{S}^o$ is the optimal solution to \eqref{D-SCM}, $V_c$ and $x^o$ are introduced in Lemma \ref{L5} and Proposition \ref{P5}, respectively.}
\end{lemma}

\begin{figure*}[t]
	\centering
	\subfigure[Time-Dependent Pricing]{\includegraphics[scale=.25]{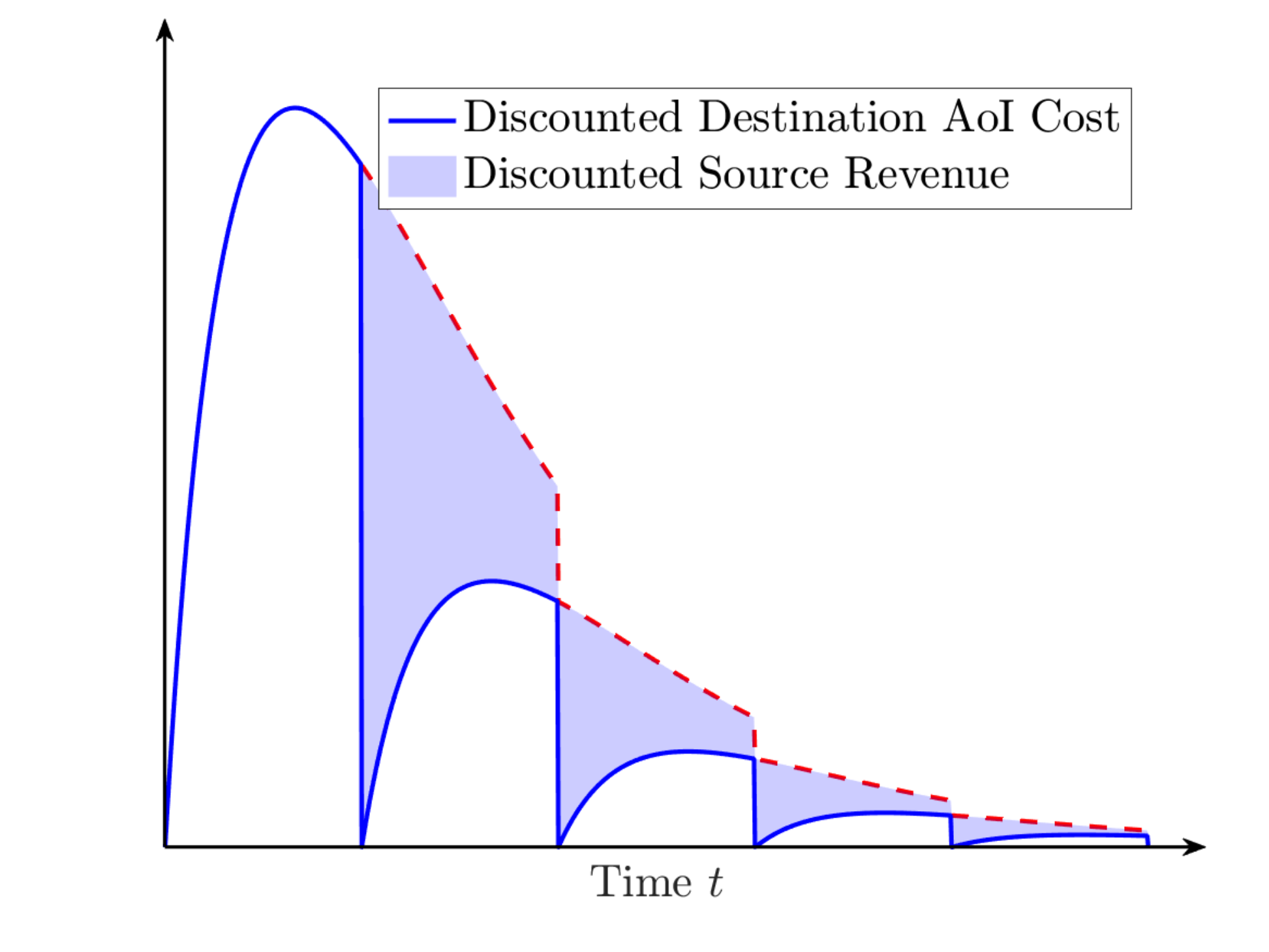}}
		\subfigure[Quantity-Based Pricing]{\includegraphics[scale=.25]{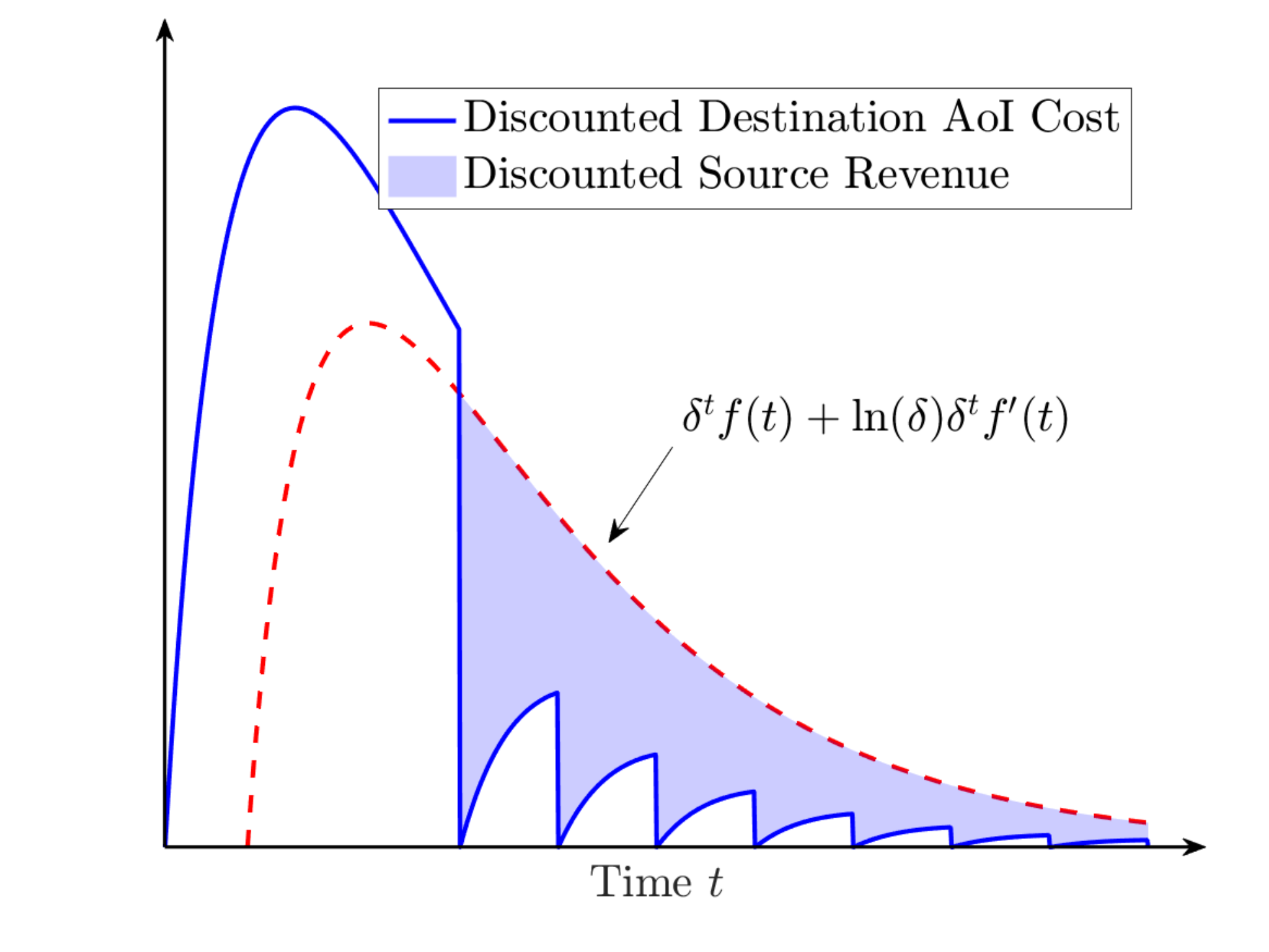}}
			\subfigure[Subscription-Based Pricing]{\includegraphics[scale=.25]{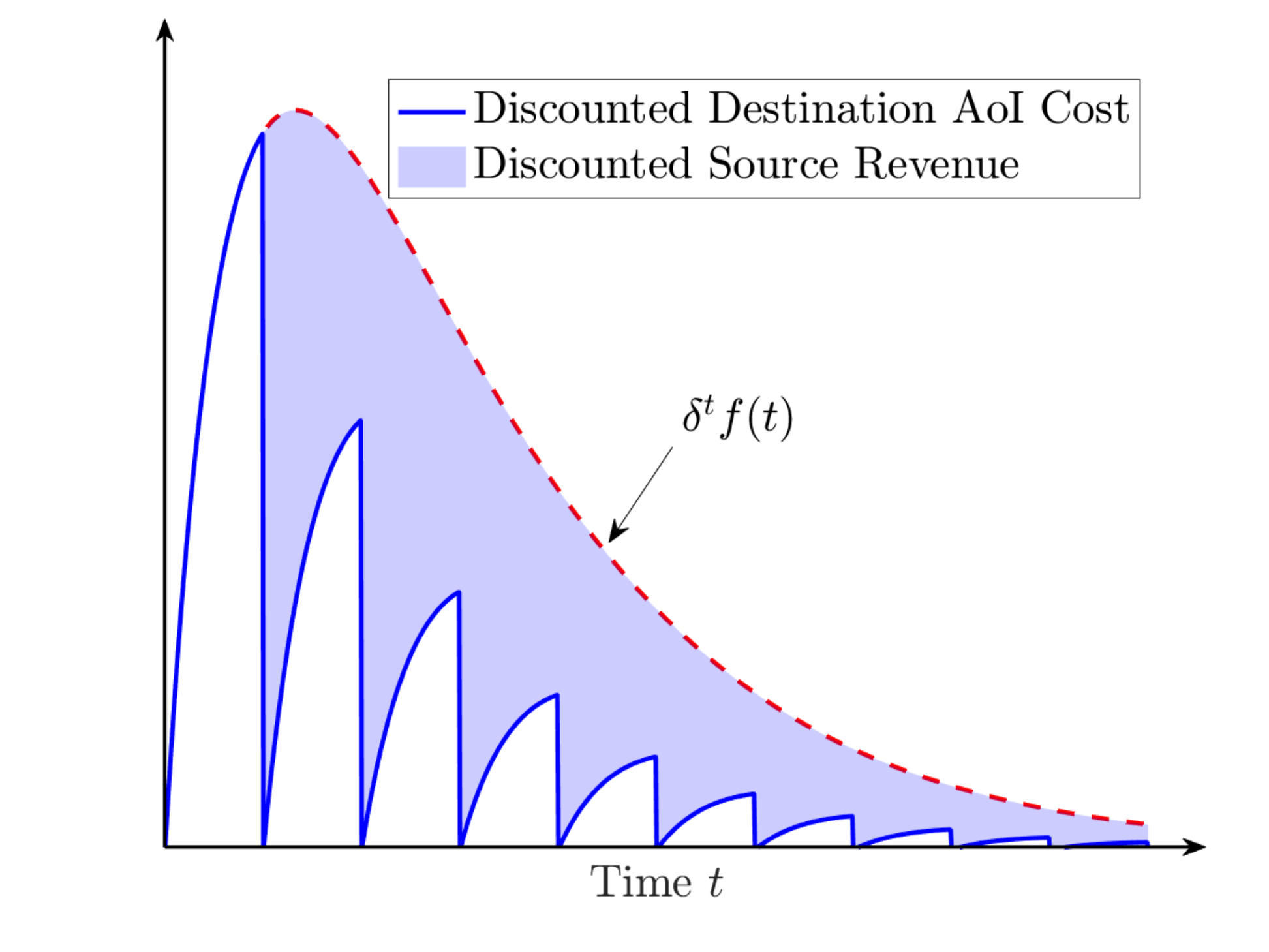}}
			\vspace{-0.2cm}
		\caption{Performance comparison in terms of the discounted AoI cost and the discounted revenue.}
		\label{Sum2}
\vspace{-0.4cm}
\end{figure*}


We present the proof of Lemma \ref{L8} in Appendix \ref{ProofL8}.
To understand Lemma \ref{L8}, the update policy after the first update (i.e., $\{S_k\}_{k\geq 2}$) is to minimize the discounted social cost. Hence, the interarrival time $S_{k+1}^{\rm \star, Q}-S_{k}^{\rm \star, Q}$ for all $k\geq 1$ is equal to $x^o$. 

Combining \eqref{asda} and Lemma \ref{L8},  we are ready to characterize the optimal quantity-based pricing:
\begin{proposition}\label{P10}
    The optimal \textit{quantity-based pricing} scheme is
\begin{align}
p_k^{\star}=\begin{cases}
\frac{1}{\ln(\delta^{-1})}(f(S_1^{\rm \star, Q})-f_\delta(x^o))-F_\delta(x^o),&~{\rm if}~k=1,\\
c(x^o),&~{\rm otherwise}.
\end{cases}\label{Eq64}
\end{align}
\end{proposition}
 
 
Intuitively,  the price after the first update is set to $c(x^o)$, ensuring that the destination's update policy in Stage II after the first update is exactly the same as the socially optimal update policy in Proposition \ref{P5}. 

		\begin{figure}[t]
		\centering
					\subfigure[]{\includegraphics[scale=.22]{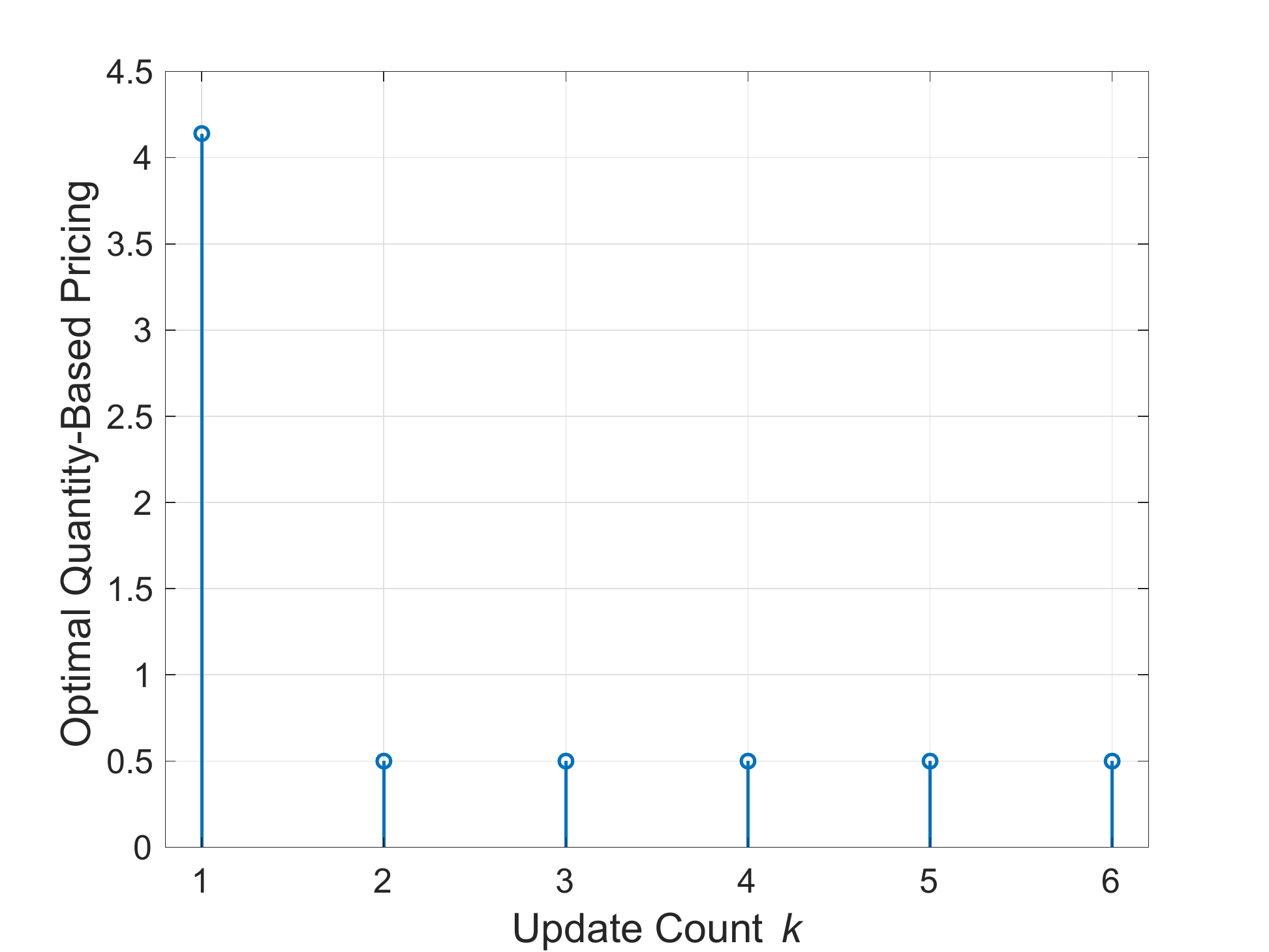}}
					\subfigure[]{\includegraphics[scale=.22]{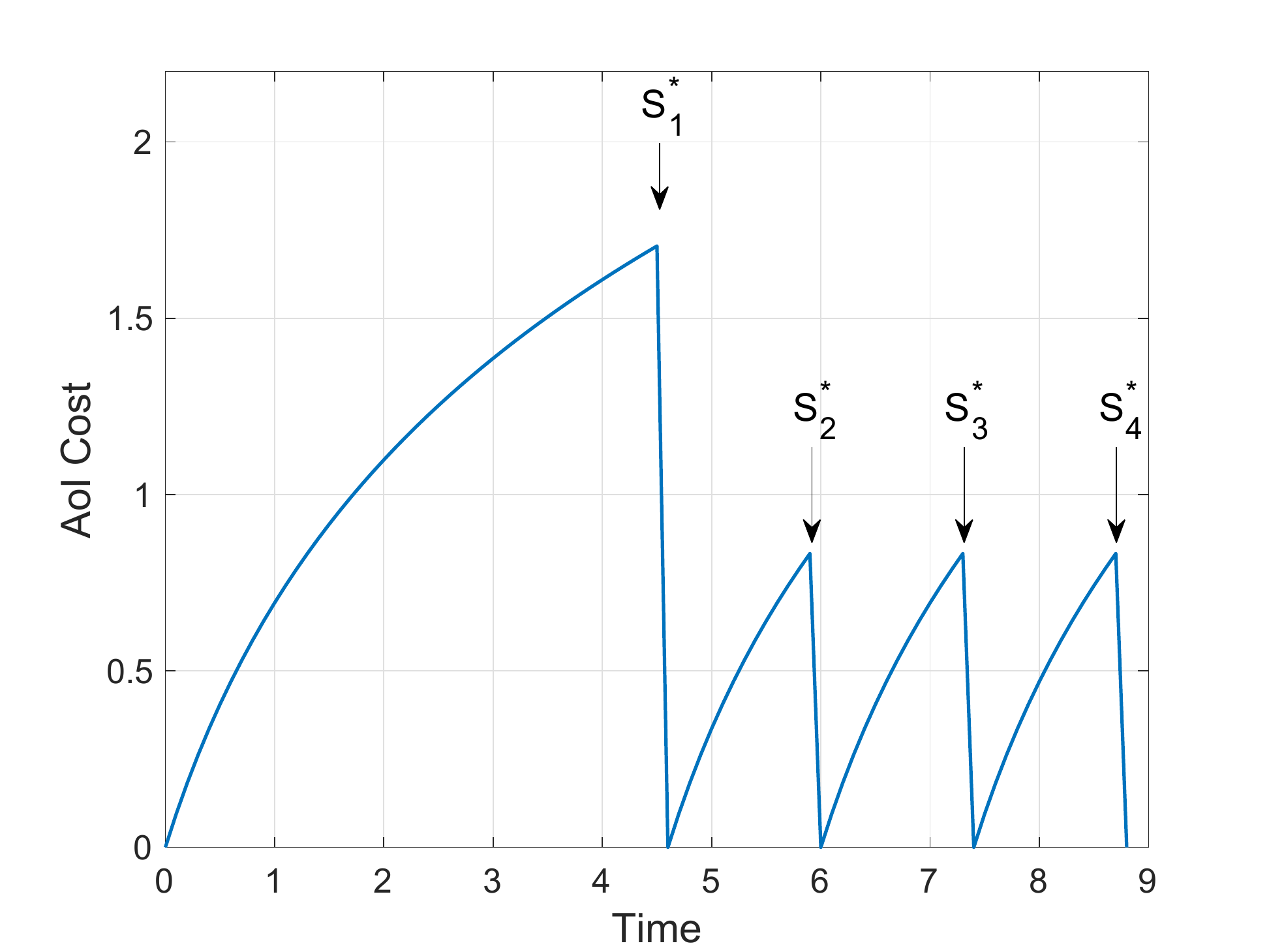}}
         \caption{An illustrative example of (a) the optimal quantity-based pricing $\Pi_q^{\star}$ and (b) the equilibrium update policy $\mathcal{S}^{\star, Q}$.}\label{PUE}
		\end{figure}
		
We illustrate the price-update equilibrium in Lemma \ref{L8} and Corollary \ref{P10} under the optimal quantity-based pricing in Fig. \ref{PUE}. As shown in Fig. \ref{PUE}(a), the source charges a relatively high price for the first update, while charging relatively low prices for the remaining updates. As illustrated in Fig. \ref{PUE}(b), such a pricing scheme leads to a larger interarrival time from \eqref{asda} compared to the later interarrival times.

\subsection{Subscription-Based Pricing Scheme}\label{SBP-U}

We finally present the subscription-based pricing $\Pi_s=\{p_u, \pi\}\in \mathbb{R}_+^2$.
{In particular, $p_u$ is the flat-rate usage price per update and is charged whenever the destination requests a data update;
$\pi$ is the subscription price and is charged at time $t=0$. }
Hence, the \textit{discounted payment} paid by the destination to the source is 
\begin{align}
    P_{\delta}(\mathcal{S},\Pi_{s})=\pi+\lim_{K\rightarrow \infty}\sum_{k=1}^{K}\delta^{S_k}p_u.
\end{align}
We note that, different from the finite-horizon model, the subscription-based pricing is not a special case of the quantity-based pricing scheme under the infinite-horizon here, since in the latter case, the source does not charge a fixed payment at $t=0$. In contrast, under the finite-horizon model, the source and the destination are insensitive to when the payment is made.

We will derive the optimal subscription-based pricing and show it is surplus-extracting, i.e., achieving the maximal source's profit among all possible pricing schemes:
\begin{proposition}\label{P-SBP-2}
	The optimal subscription-based pricing $\Pi_s^{\star}=\{p_u^{\star},\pi^{\star}\}$ is
	\begin{align}
	p_u^{\star}=c(x^o)\quad {\rm and}\quad \pi^{\star}=F_\delta(\infty)-V_c, \label{SE}
	\end{align}
	where $V_c$ and $x^o$ are introduced in Lemma \ref{L5} and Proposition \ref{P5}, respectively. In addition, $\Pi_s^\star$ is surplus-extracting (Definition \ref{Def3}).
\end{proposition}


In \eqref{SE}, $F_{\delta}(\infty)$ is the discounted aggregate cost of no data update, and $c(x^o)$ under the pricing scheme in \eqref{SE} serves to align the destination's interest to minimizing the social cost in \eqref{D-SCM}. Under the pricing in \eqref{SE}, the destination's problem becomes
	\begin{align}
&F_{\delta}(\infty)-V_c\nonumber\\
&+\min_{\mathcal{S}\in\Phi}\!\lim_{K\rightarrow \infty }\!\left[F_{\delta}(S_{1})+\sum_{k=1}^{K}\delta^{S_{k}} (F_{\delta}(S_{k+1}- S_{k})+c(\bar{x}))\right]\nonumber\\
=&F_\delta(\infty).\label{D3}
	\end{align}
\rev{The destination's discounted payoff in \eqref{D3} is $F_{\delta}(\infty)$, equal to the its discounted payoff if it does not request any update
(i.e., not subscribing to the pricing scheme).}
This
indicates that the destination will not be worse off by requesting updates (i.e., satisfying the individual rationality constraint in \eqref{IRU}). The problem in \eqref{D3} leads to the same optimal solution to the social cost minimization in \eqref{D-SCM}, and hence it corresponds to a surplus extracting pricing scheme according to Definition \ref{Def3}. Hence, from Lemma \ref{LLL2}, and the optimal subscription-based pricing in \eqref{SE} is the optimal among all possible pricing schemes.

Combining the results in Proposition \ref{P22}  and Proposition \ref{P-SBP-2}, we have the following corollary:
\begin{corollary}
	The subscription-based pricing is the optimal pricing under both finite-horizon and infinite-horizon models.
\end{corollary}

\subsection{Summary}

Finally, we summarize our key results in this section through graphically comparing the discounted AoI costs and the discounted revenues under three studied pricing schemes
 in Fig. \ref{Sum2}. \rev{Fig. \ref{Sum2}(a) presents the equal-spacing} time-dependent pricing scheme, where the discounted revenue is derived based on Lemma \ref{L6}. \rev{Fig. \ref{Sum2}(b) presents the discounted revenue of the} optimal quantity-based pricing scheme based on Lemma \ref{L66} and the fact that $\frac{1}{\ln(\delta^{-1})}f_\delta(S_1)=\int_{S_1}^\infty [f_\delta(t)+\ln(\delta) \delta^t f'(t)]dt$. In addition, the optimal quantity-based pricing (in Lemma \ref{L8} and \eqref{asda}) charges a relatively high price for the first update, and relatively low prices for the remaining updates, yielding a pricing scheme that leads to a large first interarrival time from \eqref{asda}. Finally, \rev{Fig. \ref{Sum2}(c)} presents the optimal subscription-based pricing, which is surplus-extracting (Proposition \ref{P-SBP-2}) and hence the optimal pricing scheme among all possible pricing schemes (Lemma \ref{LLL2}). The optimal subscription-based pricing induces an equal-spacing update policy as shown in Fig. \ref{Sum2}(c), which is consistent with Proposition \ref{P5}. 
Finally, comparing the discounted revenues generated by three pricing schemes in Fig. \ref{Sum2}(a)-(c), we observe that the revenue of the optimal subscription-based pricing generates more revenue than 
the suboptimal time-dependent pricing scheme and the optimal quantity-based pricing scheme do.

%
%
%
%
%

\section{Numerical Results}\label{Numerical}

In this section, we perform simulation results to compare the proposed pricing schemes. \rev{We then evaluate the significance of the performance gains of the profit-maximizing pricing, 
the impacts of time discounting, and the destination's age sensitivity on their performances.}

\subsection{Simulation Setup}
\begin{figure}[t]
	\centering
	\subfigure[]{\includegraphics[scale=.22]{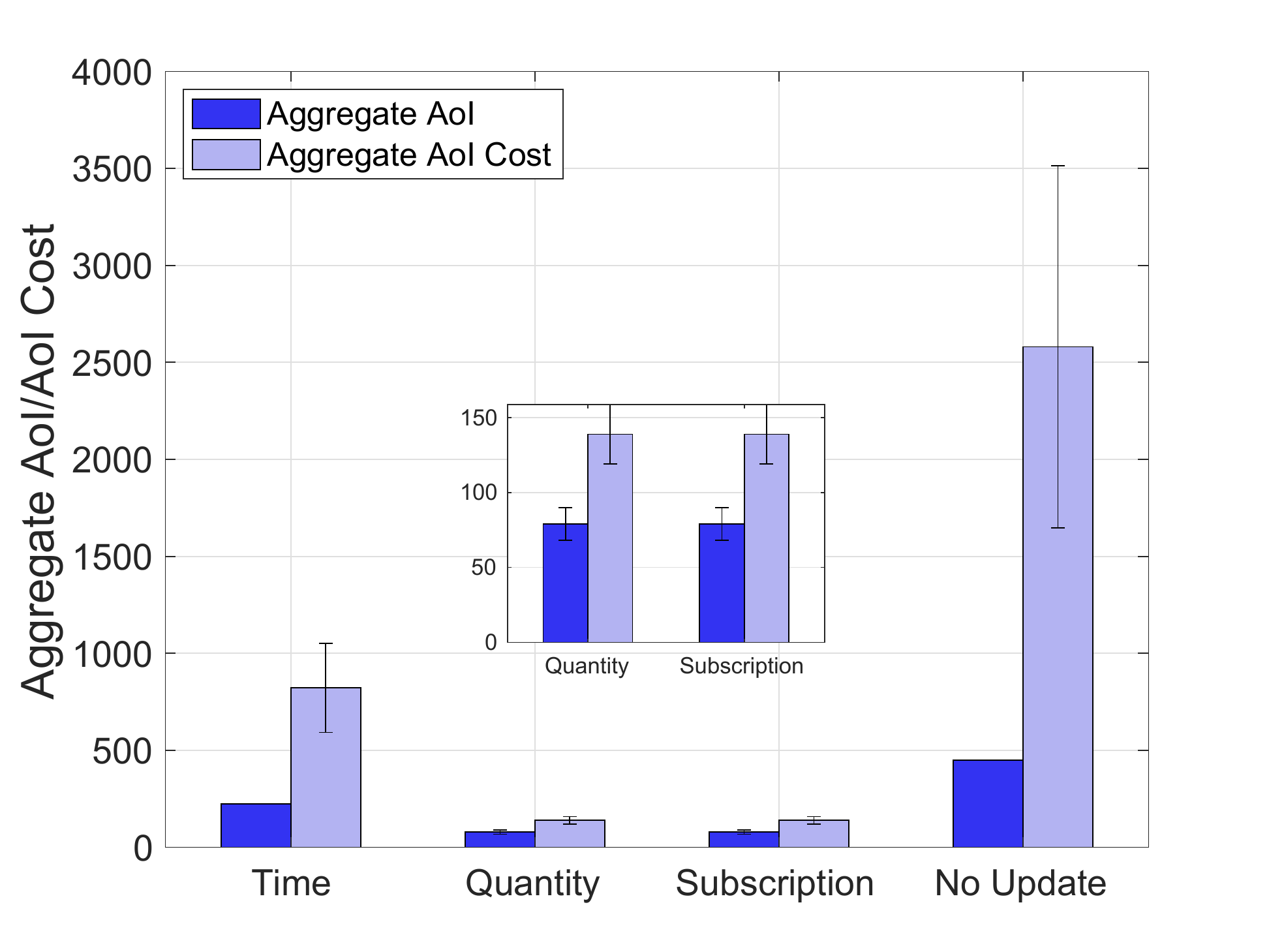}}
		\subfigure[]{\includegraphics[scale=.22]{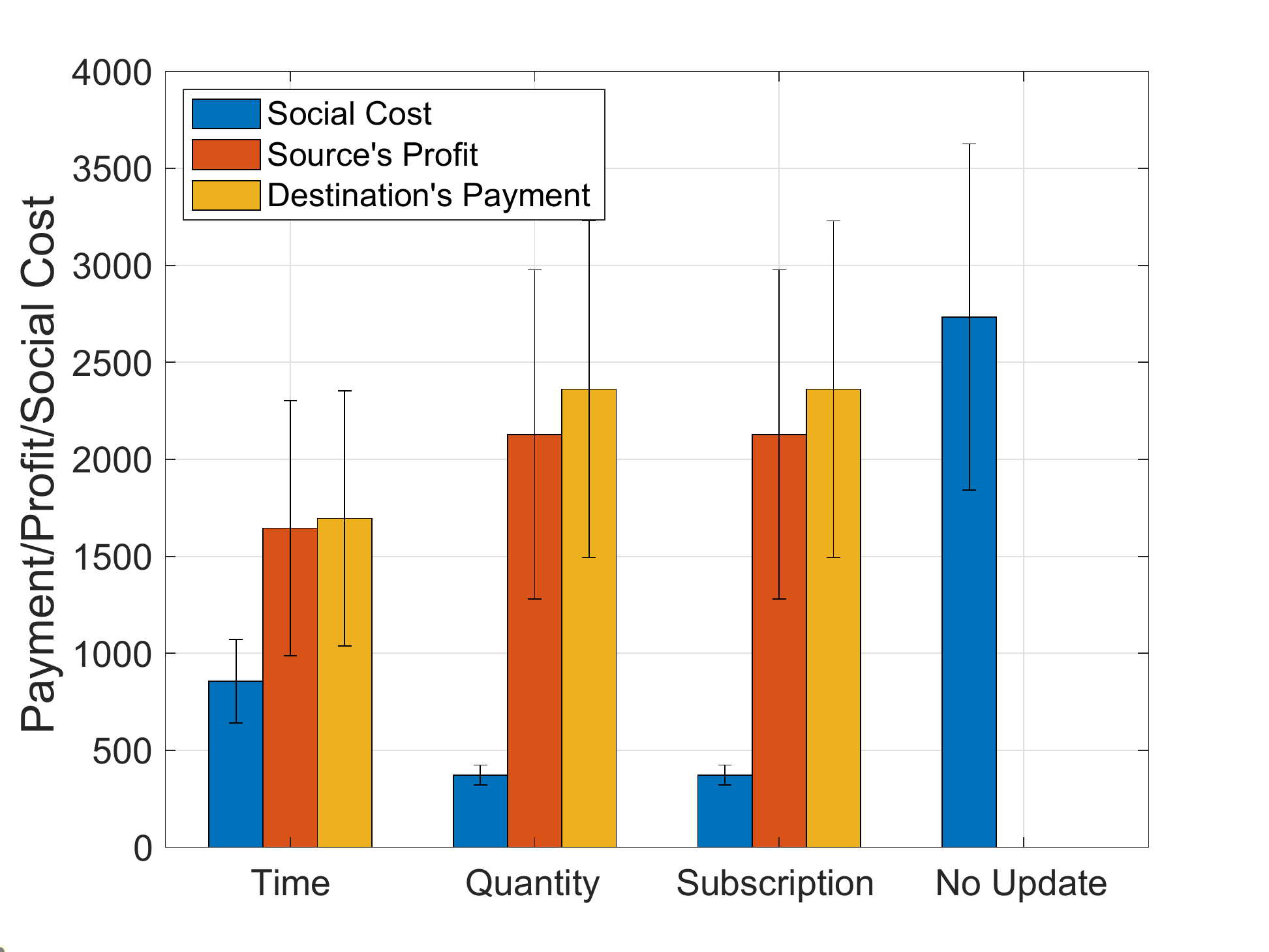}}
		\vspace{-0.2cm}
		\caption{Performance comparison in terms of (a) the discounted AoI and the discounted AoI cost, and (b) discounted profit, payment, and social cost. The error bars represent the standard deviations.}
		\label{Fig7}
 \vspace{-0.3cm}
\end{figure}

We consider a convex power AoI cost function: $f(\Delta_t)=\Delta_t^\kappa,$
where  the coefficient $\kappa\geq 1$ is termed the \textit{destination's age sensitivity.} Such an AoI cost function is useful for online learning due to the recent emergence of real-time applications such as advertisement placement and online web ranking
\cite{AoI7,online1,online2}.
Hence, the cumulative AoI cost function $F(t)$ is
$F(t)=t^{\kappa+1}/(\kappa+1)$.
The source has a constant operational cost per update, i.e., $c(\bar{x})=c$, where
$c$ is the source's \textit{operational cost coefficient}.
Let $\kappa$ follow a normal distribution $\mathcal{N}(1.5,0.2)$ truncated into the interval $[1,2]$, and let $c$ follow a normal distribution $\mathcal{N}(50,20)$ truncated into the interval $[0,100]$.  Our simulation results take the average of 100,000 experiments.

\subsection{Results for the finite-horizon Model}

\subsubsection{Performance Comparison} We compare the performances of three pricing schemes, the optimal time-dependent pricing (TDP), the optimal quantity-based pricing (QBP), and the optimal subscription-based pricing (SBP), together with a no-update (NU) benchmark. \rev{We will show that the profit-maximizing pricing schemes (the TDP and the QBP) can lead to significant profit gains compared against the benchmark (the NU).} In Fig. \ref{Fig7}(a),
we first compare the four schemes in terms of the aggregate AoI and the aggregate AoI cost. The NU scheme incurs a much larger aggregate AoI than all three proposed pricing schemes. Moreover, from Proposition \ref{P5}, the QBP and the SBP achieve the same performance, incurring an  aggregate AoI  that is  only
$59\%$ of that of the optimal TDP. 
In terms of the aggregate AoI cost, we observe a similar trend. 


In Fig. \ref{Fig7}(b), we compare the four schemes in terms of the social cost and the source's profit. We observe that  the QBP and the SBP are $27\%$ more profitable than the TDP. 
In addition, the optimal TDP only incurs $34\%$ of the social cost of the NU scheme. The optimal QBP and SBP further reduce the social cost and incur only $46\%$ of that of the optimal TDP. 
Therefore, \rev{\textit{the profit-maximizing pricing schemes (the TDP and the QBP) can significantly outperform the benchmark in terms of the aggregate AoI cost and the social cost.}}

\begin{figure}[t]
	\centering
	\subfigure[]{	
			\includegraphics[scale=0.21]{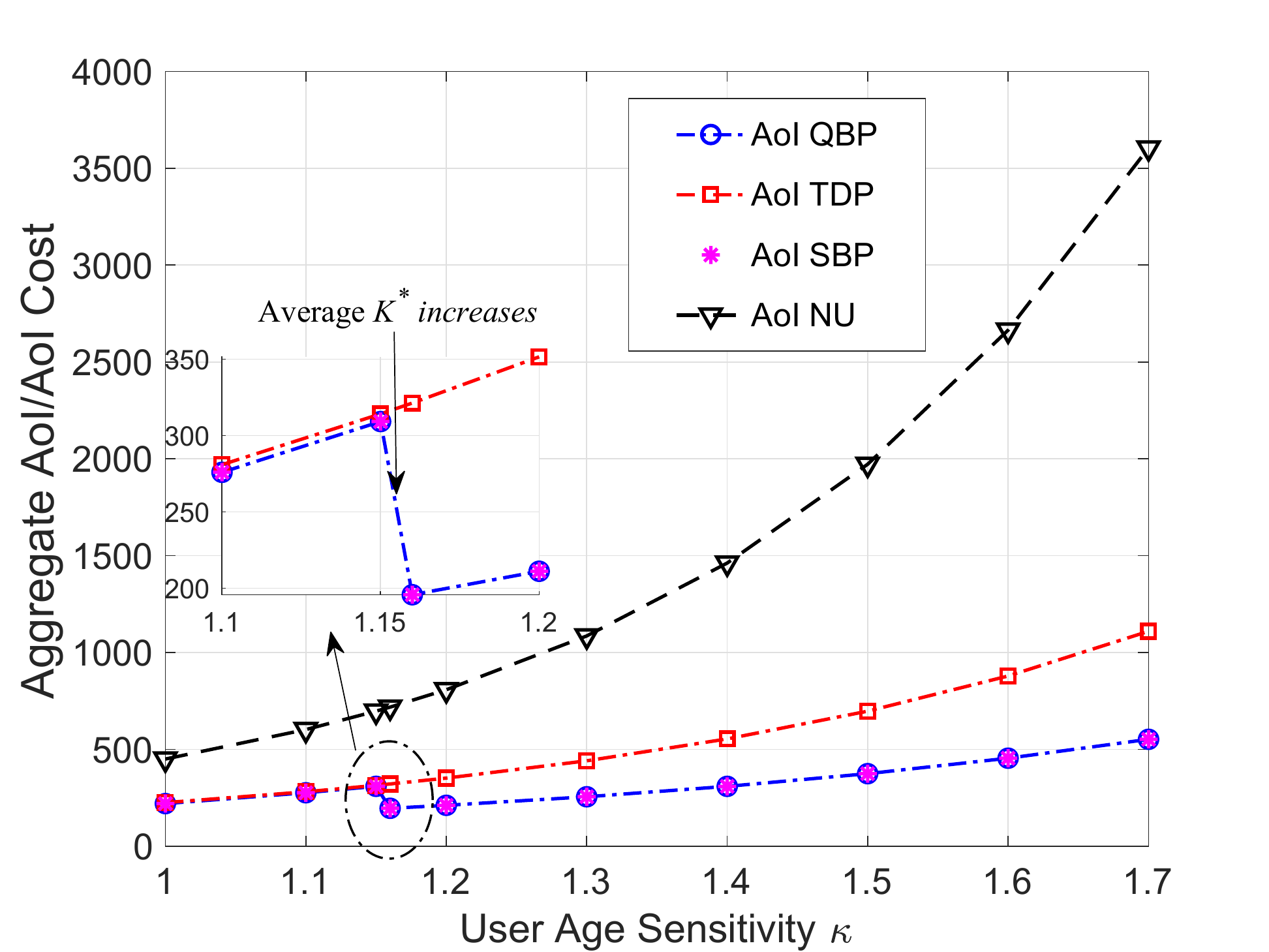}}
	\subfigure[]{	
			\includegraphics[scale=0.21]{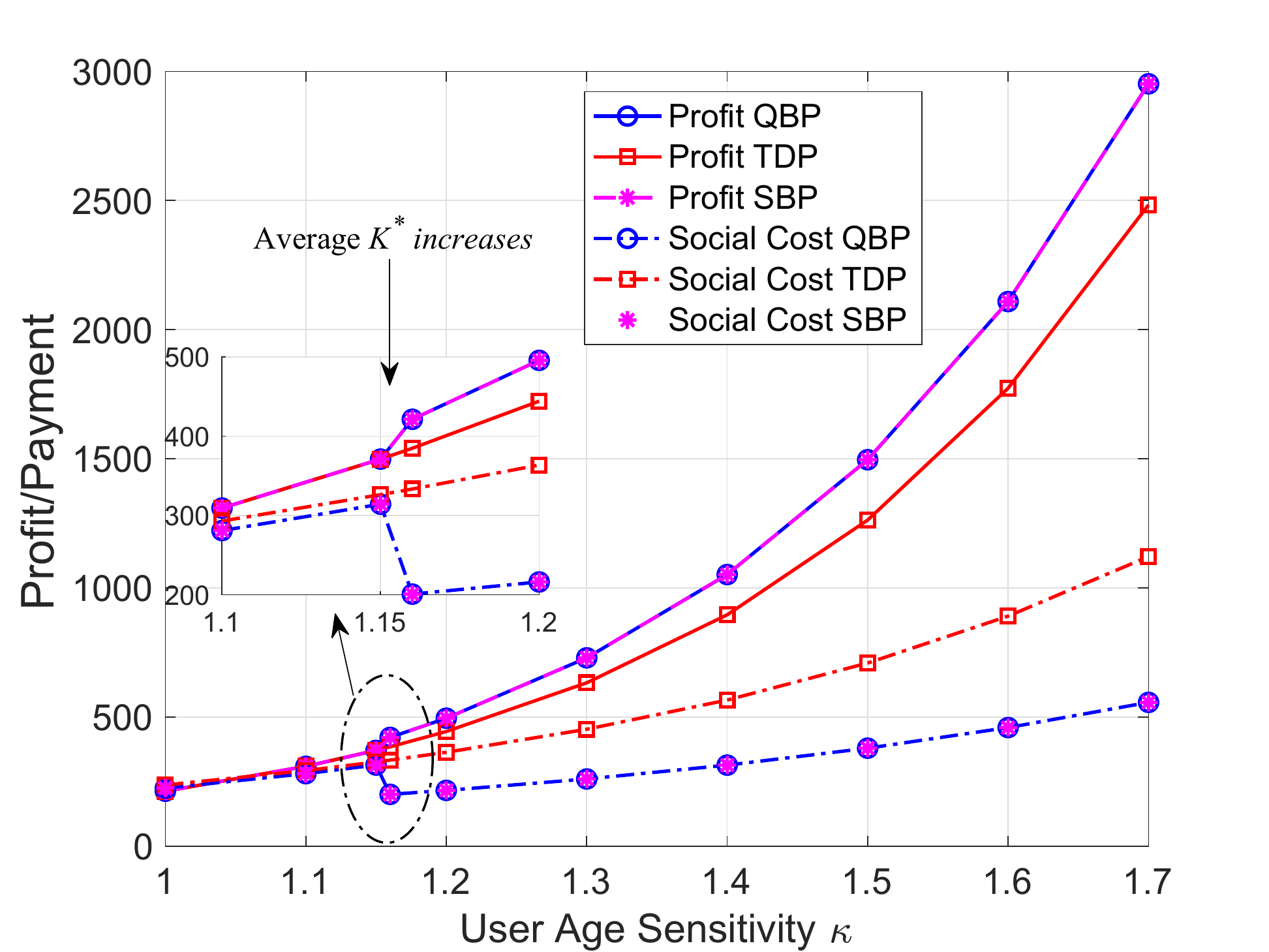}}
			\vspace{-0.2cm}
		\caption{Impact of destination's age sensitivity $\kappa$ on the aggregate AoI and the AoI cost.}
		\label{Fig11}
 \vspace{-0.4cm}
\end{figure}
\subsubsection{Impact of Age Sensitivity} Fig. \ref{Fig11}(a) compares the performances of the four schemes at different age sensitivities $\kappa$, which characterizes how the destination is sensitive to the AoI. First, the QB*, The TDP, and the SBP lead to the same aggregate AoI under small $\kappa<1.16$. This is because   the TDP scheme always leads to one update while a small age sensitivity also renders a small amount of total updates for the QBP and the SBP.
Second, when $\kappa$ is increased to $1.16$, there is a small decrease in aggregate AoI for the QBP and SBP schemes. This is due to the fact that $\kappa$ increases the number of updates $K^*$, as the destination becomes more sensitive to the AoI. 
Third, as $\kappa$ increases, we see that the AoI cost increases for both the NU scheme and the TDP. However, the aggregate AoI cost for the TDP increases much slower than the NU scheme while the AoI cost for the QBP and SBP schemes increases even slower.
We observe a similar trend for the payment ga*, The profit gap, and  the social cost gap between the TDP and the QBP (SBP); they increase as $\kappa$ increases in Fig. \ref{Fig11}(b). That is, \emph{the destination's sensitivity to the age increases the performance gaps between the optimal pricing schemes (the QBP and the SBP) and the benchmark (the NU)}.

\subsection{Results for the Infinite-Horizon Model}

\begin{figure}[t]
				\centering
		\subfigure[]{\includegraphics[scale=.22]{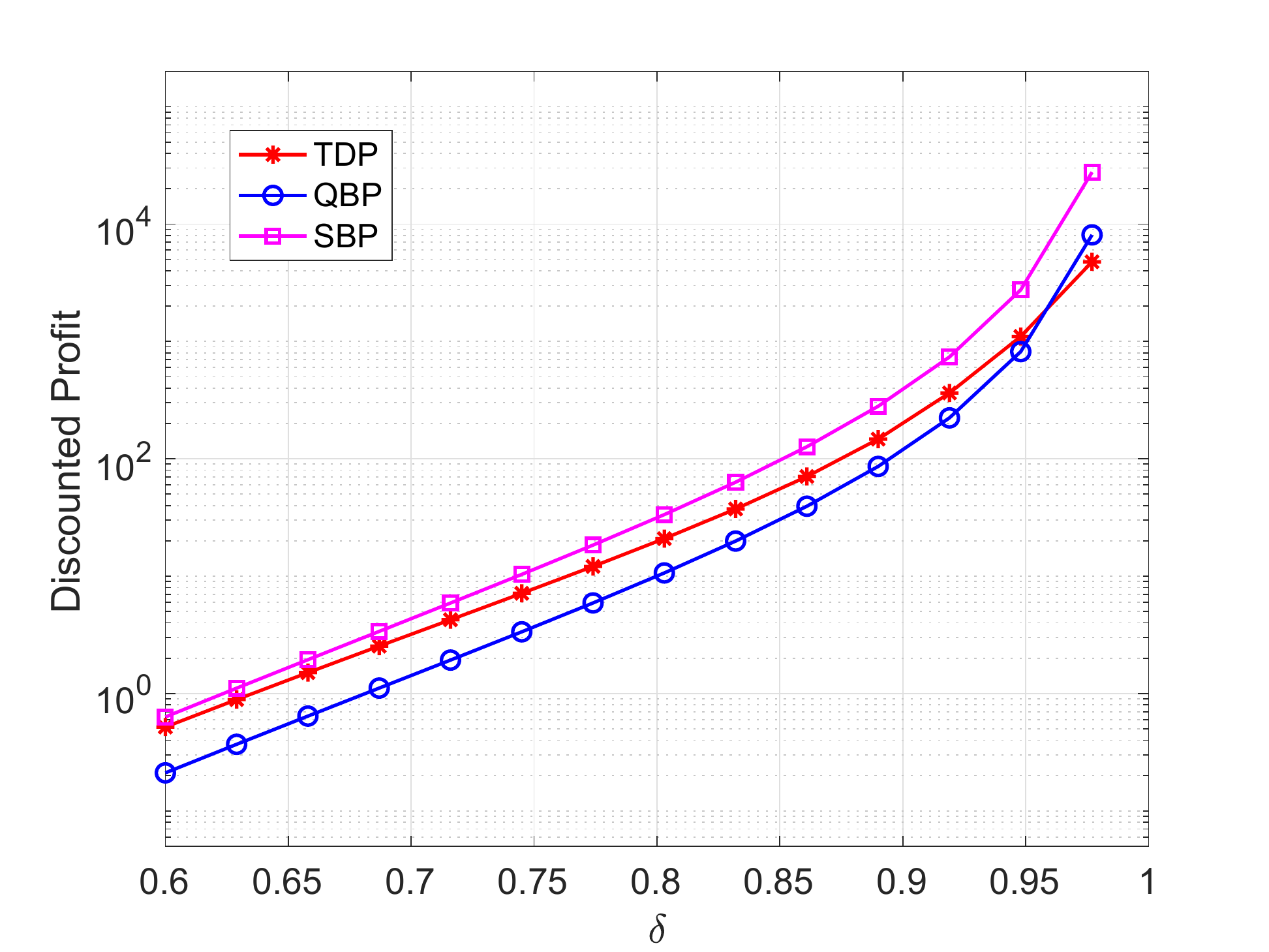}}
		\subfigure[]{\includegraphics[scale=.22]{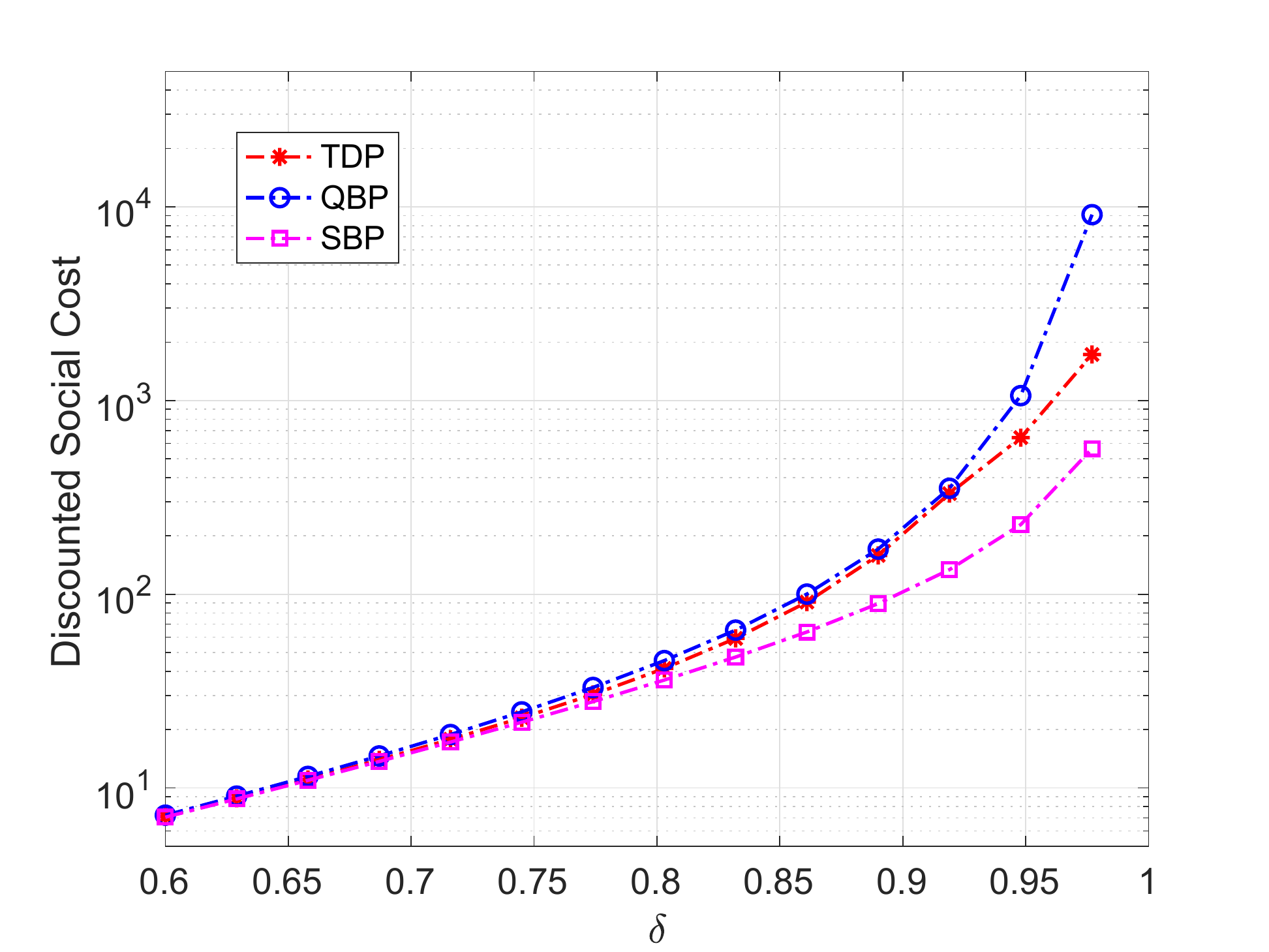}}
			\vspace{-0.1cm}
		\caption{Impacts of the discounted coefficient $\delta$ on (a) discounted profit and (b) discounted social cost.}\label{Z124}
		\vspace{-0.1cm}
		\label{f3a}
	\end{figure}

 We now present numerical results for the infinite-horizon model.
Fig. \ref{Z124} compares the performances of the three pricing schemes under the discount coefficient $\delta$ to demonstrate how the time discounting affects the performances of different pricing schemes. In Fig. \ref{Z124}(a), we observe that the optimal SBP is more profitable than
the optimal QBP and the suboptimal TDP, as the optimal SBP is the optimal pricing scheme among all possible pricing schemes (Proposition \ref{P-SBP-2}). An interesting observation is that the TDP outperforms the QBP when $\delta<0.97$, and the QBP outperforms the TDP when $\delta$ is large. Hence, different from the finite-horizon model, the QBP does not always perform better than the TDP due to the time discounting. Moreover, as $\delta$ decreases, the performance of the TDP performs closely to the SBP, which is consistent with the result in Proposition \ref{Asy} that the TDP is asymptotically surplus-extracting. More importantly, as $\delta$ approaches $0.6$, the TDP  already performs very close to the SBP. This
implies that \textit{a moderate degree of time discounting is enough to make the TDP close-to-profit-maximizing. }





In Fig. \ref{Z124}(b), we observe that the SBP achieves a smaller social cost compared against
the QBP and the TDP, as the surplus-extracting pricing scheme also achieves the minimal social cost by Definition \ref{Def3}.
 An interesting observation is that all three pricing schemes 
 perform more closely to each other as $\delta$ decreases and achieve the discounted social cost with negligible differences
 under a moderate level of time discounting $\delta$ (i.e., $\delta=0.6$).

\section{Conclusions}\label{Conclusion}

We presented the first pricing scheme design for fresh data trading and proposed three pricing schemes to explore the profitability of exploiting different dimensions in designing pricing.
Our results revealed that (i) the profitability to exploit the time flexibility depends on the degree of time discounting;
(ii) the optimal quantity-based pricing scheme achieves the maximal source's profit among all pricing schemes with a  finite-horizon model but not with an infinite-horizon model; (iii) the optimal low-complexity subscription-based pricing scheme achieves the maximal source's profit under both models.

Our results shed light on pricing scheme design for a more general scenario: multi-destination systems, which raise the challenges of
coupling system constraints (e.g., interference constraints). Another interesting direction is to study incomplete information settings, which requires leveraging \textit{mechanism design} to elicit destinations' truthful information regarding AoI.

\begin{IEEEbiography}[{\includegraphics[width=1in,height=1.25in,clip,keepaspectratio]{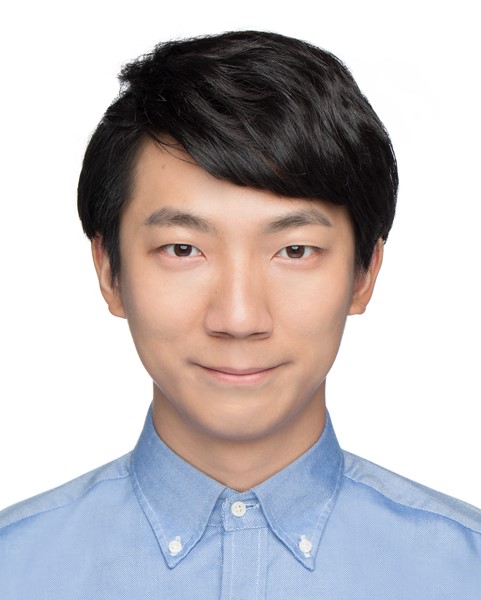}}]{Meng Zhang} (S'15, M'19) is currently a Postdoctoral Fellow with the Department of Electrical and Computer Engineering at Northwestern University. He received his Ph.D. degree in Information Engineering from the Chinese University of Hong Kong in 2019. He was a visiting student research collaborator with the Department of Electrical Engineering at Princeton University  from 2018 to 2019. His primary research interests include network economics and wireless networks, with a current emphasis on mechanism design and optimization for age-of-information.
\end{IEEEbiography}

\begin{IEEEbiography}[{\includegraphics[width=1in,height=1.25in,clip,keepaspectratio]{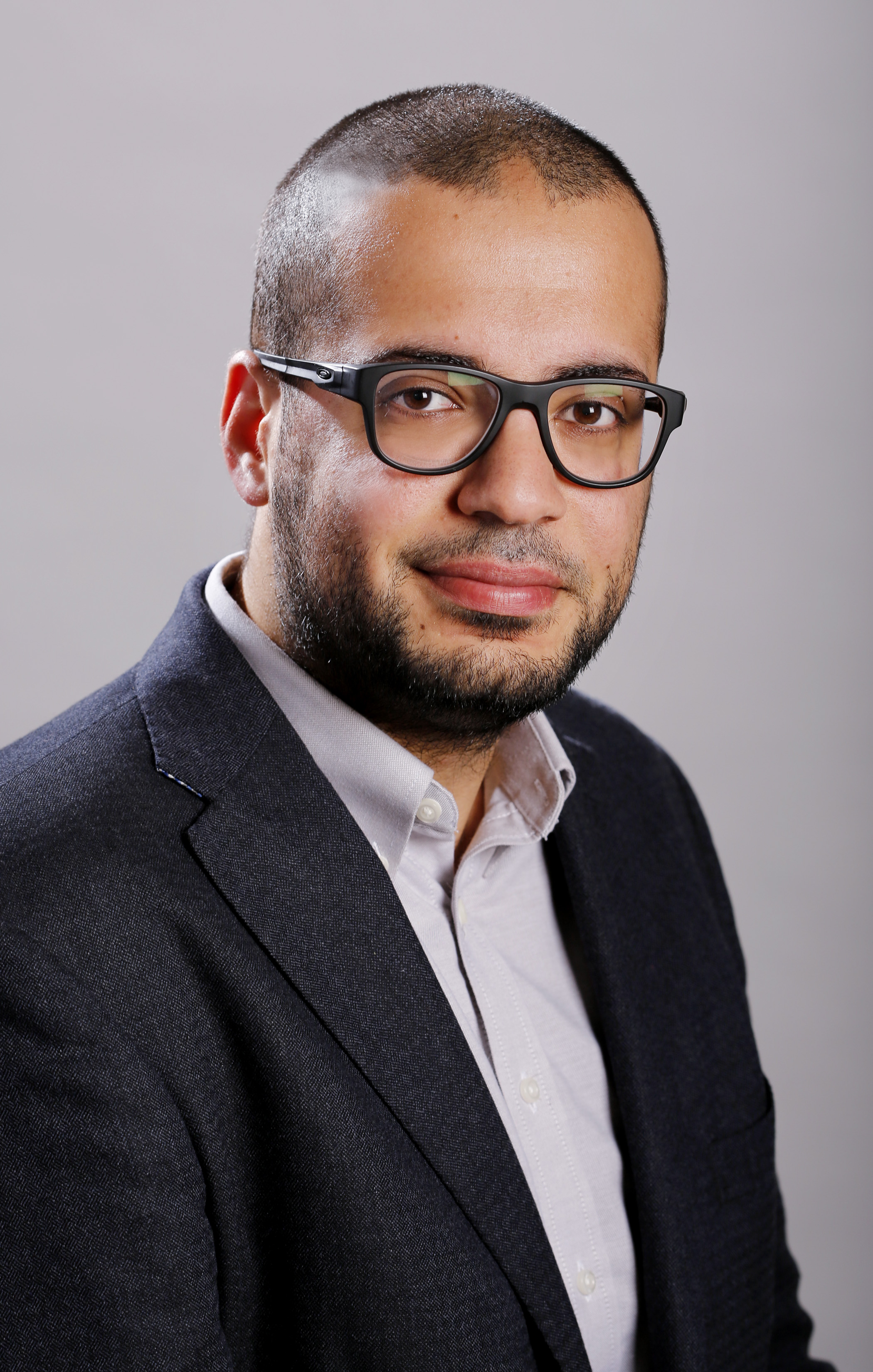}}]{Ahmed Arafa} (S'13, M'17) received the B.Sc.~degree, with highest honors, in electrical engineering from Alexandria University, Egypt, in 2010, the M.Sc.~degree in wireless technologies from the Wireless Intelligent Networks Center (WINC), Nile University, Egypt, in 2012, and the M.Sc.~and Ph.D.~degrees in electrical engineering from the University of Maryland at College Park, MD, USA, in 2016 and 2017, respectively. He has been with the Electrical Engineering Department at Princeton University as a Postdoctoral Research Associate during 2017--2019. Currently, he is an Assistant Professor in the Department of Electrical and Computer Engineering at the University of North Carolina at Charlotte.

Dr.~Arafa's research interests are in communication theory, information theory, machine learning, and signal processing, with recent focus on timely information processing and transfer (age-of-information), energy harvesting communications, information-theoretic security and privacy, and federated learning. He was the recipient of the Distinguished Dissertation award from the Department of Electrical and Computer Engineering, University of Maryland, in 2017, for his Ph.D.~thesis work on optimal energy management policies in energy harvesting communication networks with system costs.
\end{IEEEbiography}

\begin{IEEEbiography}[{\includegraphics[width=1in,height=1.25in,clip,keepaspectratio]{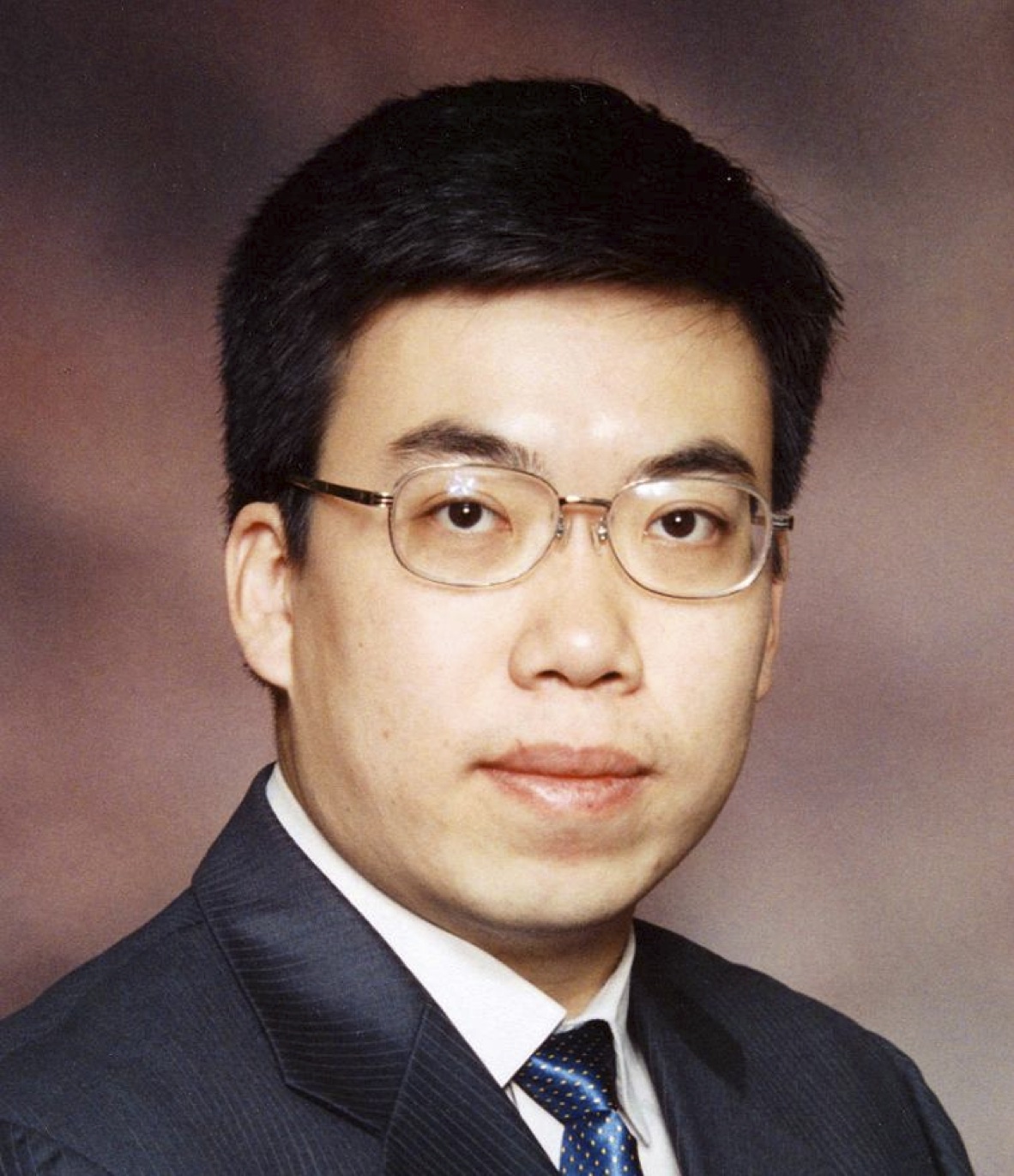}}]{Jianwei Huang}
(F'16) received the Ph.D. degree in ECE from Northwestern University in 2005, and worked as a Postdoc Research Associate in Princeton University during 2005-2007. From 2007 until 2018, he was on the faculty of Department of Information Engineering, The Chinese University of Hong Kong. Since 2019, he has been on the faulty at The Chinese University of Hong Kong, Shenzhen, where he is currently a Presidential Chair Professor and an Associate Dean of the School of Science and Engineering. He also serves as a Vice President of Shenzhen Institute of Artificial Intelligence and Robotics for Society. His research interests are in the area of network optimization, network economics, and network science, with applications in communication networks, energy networks, data markets, crowd intelligence, and related fields. He has co-authored 9 Best Paper Awards, including the 2011 IEEE Marconi Prize Paper Award in Wireless Communications. He has co-authored seven books, including the textbook on "Wireless Network Pricing.” He is an IEEE Fellow, and was an IEEE ComSoc Distinguished Lecturer and a Clarivate Web of Science Highly Cited Researcher. He is the Editor-in-Chief of IEEE Transactions on Network Science and Engineering, and was the Associate Editor-in-Chief of IEEE Open Journal of the Communications Society. 
\end{IEEEbiography}

\begin{IEEEbiography}[{\includegraphics[width=1in,height=1.25in,clip,keepaspectratio]{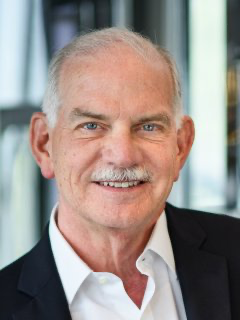}}]{H. Vincent Poor} (S’72, M’77, SM’82, F’87) received the Ph.D. degree in EECS from Princeton University in 1977.  From 1977 until 1990, he was on the faculty of the University of Illinois at Urbana-Champaign. Since 1990 he has been on the faculty at Princeton, where he is currently the Michael Henry Strater University Professor. During 2006 to 2016, he served as the dean of Princeton’s School of Engineering and Applied Science. He has also held visiting appointments at several other universities, including most recently at Berkeley and Cambridge. His research interests are in the areas of information theory, machine learning and network science, and their applications in wireless networks, energy systems and related fields. Among his publications in these areas is the forthcoming book \textit{Machine Learning and Wireless Communications.}  (Cambridge University Press, 2021).

Dr. Poor is a member of the National Academy of Engineering and the National Academy of Sciences and is a foreign member of the Chinese Academy of Sciences, the Royal Society, and other national and international academies. Recent recognition of his work includes the 2017 IEEE Alexander Graham Bell Medal and a D. Eng. \textit{honoris causa} from the University of Waterloo awarded in 2019.	
\end{IEEEbiography}	
	
\newpage	
	
\appendix

\section{Proofs}

\subsection{Proof of Lemma \ref{LLL1}}\label{ProofLLL1}
By the individual rationality constraint in \eqref{IRP}, all pricing schemes need to satisfy
\begin{align}
P(\mathcal{S}^o(\Pi),\Pi)\leq F(T)-\Gamma(\mathcal{S}^o(\Pi)).
\end{align}
Hence, the source's profit thus is
\begin{align}
P(\mathcal{S}^o(\Pi))-C(K^o(\Pi))\leq &~F(T)-\Gamma(\mathcal{S}^o(\Pi))-C(K^o(\Pi)),\nonumber\\
\leq &~F(T)-\min_{\mathcal{S}\in\Phi}[\Gamma(\mathcal{S})+C(K)].\label{eq4444}
\end{align}
Hence, if a pricing scheme achieves the upper bound in \eqref{eq4444}, it achieves the maximal profit among all pricing schemes.

 
\subsection{Proof of Lemma \ref{L1}}\label{ProofL1}



	Given an equilibrium $(\Pi_t^*,K^{\rm *,T},\bs{x}^{\rm *,T})$, we first prove that $p^{*}\left(\sum_{j=1}^kx_j^{\rm *,T}\right)\leq 
	DF(x_{k+1}^{\rm *,T},x_k^{\rm *,T})$ for all $ k\in\mathcal{K}(K^{\rm *,T}+1)$. The equilibrium destination's overall cost (the aggregate AoI cost plus the payment) is
	\begin{align}\label{D1}
	\sum_{k=1}^{K^{\rm *,T}}F(x_k^{\rm *,T})+\sum_{k=1}^{K^{\rm *,T}+1}p^{\rm *}\left(\sum_{j=1}^kx_j^{\rm *,T}\right).
	\end{align}
	Suppose that there exists an update $l$ such that $p^{\rm *}\left(\sum_{j=1}^lx_j^{\rm *,T}\right)>
	DF(x_{l+1}^{\rm *,T},x_l^{\rm *,T})$. Suppose that the destination changes its update policy from $\mathcal{S}^{\rm *,T}$ to $\mathcal{S}^{\rm *,T}/\{S_l^{\rm *,T}\}$
	(i.e., removing the update at $S_l^{\rm *,T}$), the overall cost becomes 
	\begin{align}\label{D22}
	\sum_{k=1,k\notin \{l,l+1\}}^{K^{\rm *,T}}\!\!\!\! F\left(x_k^{\rm *,T}\right)+F(x^{\rm *,T}_l+x^{\rm *,T}_{l+1})+\!\!\!\!\sum_{k=1,k\neq l}^{K^{\rm *,T}+1}\!\!\!\!p^{\rm *}\left(\sum_{j=1}^kx_j^{\rm *,T}\right).
	\end{align}	
	Subtracting \eqref{D22} by \eqref{D1}, we have
	\begin{align}
	&F(x^{\rm *,T}_l+x^{\rm *,T}_{l+1})-F(x_l^{\rm *,T})-F(x_{l+1}^{\rm *,T})-p^{\rm *}\left(\sum_{j=1}^lx_j^{\rm *,T}\right)\nonumber\\
	=&DF(x^{\rm *,T}_{l+1},x^{\rm *,T}_l)-p^{\rm *}\left(\sum_{j=1}^lx_j^{\rm *,T}\right)<0.
	\end{align}
	This means that by removing the update at $S_l^{\rm *,T}$, the destination can reduce its overall cost, contradicting with the fact that $\bs{x}^{\rm *,T}$ is at an equilibrium. Hence, we prove that  $p^{\rm *}\left(\sum_{j=1}^kx_j^{\rm *,T}\right)\leq 
	DF(x_{k+1}^{\rm *,T},x_k^{\rm *,T})$ for all $ k\in\mathcal{K}(K^{\rm *,T}+1)$ at the equilibrium.
	
	To prove $p^{\rm *}\left(\sum_{j=1}^kx_j^{\rm *,T}\right)\geq
	DF(x_{k+1}^{\rm *,T},x_k^{\rm *,T})$ for all $ k\in\mathcal{K}(K^{\rm *,T}+1)$, we adopt the following lemma:
	 \begin{lemma}\label{L33}
	 For any time-dependent pricing scheme $p(t)$ leading to the destination's update policy $\mathcal{S}^{\rm *,T}(p(t))$, 
	we can always construct another time-dependent pricing scheme $\tilde{p}(t)$ given by
	\begin{align}\label{inftyprice}
	\tilde{p}(t)=\begin{cases}
	p(t),&~{\rm if}~t=S_k,~\forall S_k\in\mathcal{S}^{\rm *}(p(t)),\\
	+\infty,&~{\rm otherwise},
	\end{cases}
	\end{align}
	which leads to the same destination's update policy, i.e., $\mathcal{S}^{\rm *}(\tilde{p}(t))=\mathcal{S}^{\rm *}(p(t))$, and hence the same source's profit. 
	\end{lemma}
\begin{IEEEproof}
	To see why both pricing scheme lead to the same destination's update policy, suppose that $\mathcal{S}^{\rm *}(p(t))\neq \mathcal{S}^{\rm *}(\tilde{p}(t))$. We have that  (i) under $\tilde{p}(t)$, the update policy  $\mathcal{S}^{\rm *}(\tilde{p}(t))$ leads to a smaller destination's overall cost; (ii) $\mathcal{S}^{\rm *}(\tilde{p}(t))\subset \mathcal{S}^{\rm *}(p(t))$, since the infinitively large prices in \eqref{inftyprice} make the destination unable to update at the time instances outside the set $\mathcal{S}^{\rm *}(p(t))$. Hence, under both pricing schemes $p(t)$ and $\tilde{p}(t)$, $\mathcal{S}^{\rm *}(\tilde{p}(t))$ leads to the same destination's overall cost which is smaller than the overall cost incurred by $\mathcal{S}^{\rm *}(p(t))$. This contradicts with the fact that $\mathcal{S}^{\rm *}(p(t))$ is the optimal update policy under $p(t)$. Therefore, we must have $\mathcal{S}^{\rm *}(\tilde{p}(t))=\mathcal{S}^{\rm *}(p(t))$.
\end{IEEEproof}

	We then prove $p^{\rm *}\left(\sum_{j=1}^kx_j^{\rm *,T}\right)\geq
	DF\left(x_{k+1}^{\rm *,T},x_k^{\rm *,T}\right)$ for all $ k\in\mathcal{K}(K^{\rm *,T}+1)$ by considering the following lemma:
	\begin{lemma}\label{L444}
	Consider a time-dependent pricing such that
	\begin{align}
{p}(t)\begin{cases}
\leq DF(\tilde{x}_{k+1},\tilde{x}_k),&~{\rm if}~t=\tilde{S}_k,~\forall \tilde{S}_k\in\tilde{\mathcal{S}},\\
=+\infty,&~{\rm otherwise},\label{ppp}
\end{cases}
\end{align}
for some update policy $\tilde{\mathcal{S}}=\{\tilde{S}_k\}$. We have $\mathcal{S}^{\rm *}(p(t))=\tilde{\mathcal{S}}$.
	\end{lemma}
\begin{IEEEproof}
We prove the Lemma \ref{L444} by induction:
		\begin{itemize}
	\item \textbf{Base case:}  Regardless of the destination's all previous update candidate time instances $\{\tilde{S}_1,\tilde{S}_2,...,\tilde{S}_{K-1}\}$, the minimal AoI at the time instance $\tilde{S}_K$ is $\tilde{x}_{K}$. In this case, the minimal aggregate AoI cost reduction is at $DF(\tilde{x}_{K+1},\tilde{x}_K)$. Hence, when $p(\tilde{S}_K)\leq  DF(\tilde{x}_{K+1},\tilde{x}_K)$, the destination would update at $\tilde{S}_K$.
	\item \textbf{Induction Step:} Suppose that the destination would update at $\{\tilde{S}_{j}\}_{k+1\leq j\leq K}$. Regardless of the destination's all previous update candidate time instances $\{\tilde{S}_{j}\}_{1\leq j\leq k-1}$, the minimal AoI at the time instance $\tilde{S}_k$ is $\tilde{x}_{k}$ due to the pricing scheme in \eqref{ppp}. In this case, the minimal aggregate AoI cost reduction is at $DF(\tilde{x}_{k+1},\tilde{x}_k,0)$.  Hence, when $p(\tilde{S}_k)\leq  DF(\tilde{x}_{k+1},\tilde{x}_k,0)$, the destination would update at $\tilde{S}_k$, since it can reduce the overall cost (the aggregate AoI cost plus the payment).
	\end{itemize}
Hence, by the principle of induction, we complete the proof of Lemma \ref{L444}.
\end{IEEEproof}

According to Lemma \ref{L444}, if there exists an equilibrium where the pricing scheme is  $\tilde{p}^{\rm *}\left(\sum_{j=1}^kx_j^{\rm *, T}\right) < DF(x_{k+1}^{\rm *, T},x_k^{\rm *, T})$, the source can always increase $\tilde{p}^{\rm *}\left(\sum_{j=1}^kx_j^{\rm *, T}\right)$ to $ DF(x_{k+1}^{\rm *, T},x_k^{\rm *, T})$ to
	improve its profit without changing the destination's update policy, and hence improves its profit. 
	This contradicts with the fact that $(p^{\rm *}(t),K^{\rm *, T},\bs{x}^{\rm *, T})$ is an equilibrium and hence we have $p^{\rm *}\left(\sum_{j=1}^kx_j^{\rm *, T}\right)\geq
	DF(x_{k+1}^{\rm *, T},x_k^{\rm *, T})$ for all $ k\in\mathcal{K}(K^{\rm *, T}+1)$.
	
	Combining the above discussions of two cases, we complete the proof.

\subsection{Proof of Proposition \ref{P1}}\label{ProofP1}

Lemma \ref{L1} implies that the destination's payment is 
$$\sum_{k=1}^{K^{\rm *, T}}DF(x_{k+1}^{\rm *, T},x_k^{\rm *, T}).
$$
and the source's achievable profit is
\begin{align}
\sum_{k=1}^{K^{\rm *, T}}DF(x_{k+1}^{\rm *, T},x_k^{\rm *, T})-C(K^{\rm *, T}),
\end{align}
for some equilibrium $(K^{\rm *, T},\bs{x}^{\rm *, T})$. Hence, the optimal value of problem \eqref{Refor} leads to  the maximal achievable profit for the source.

We next show that, given the optimal solution $(K^o,\bs{x}^o)$  to problem \eqref{Refor}, the following  is an optimal time-dependent pricing scheme:
		\begin{align}
p^{\rm *}\left(t\right)=\begin{cases}
DF(x_{k+1}^{\rm *, T},x_k^{\rm *, T}),~&{\rm if}~t=\sum_{j=1}^kx_j^{\rm *, T},~\forall k\in\mathcal{K}(K^o+1),\\
+\infty,~&{\rm otherwise}.\end{cases}\label{zzzz}
\end{align}
From Lemma \ref{L444}, the under the pricing in \eqref{zzzz}, the destination's update policy is $(K^{\rm *, T},\bs{x}^{\rm *, T})=(K^{\rm *, T},\bs{x}^{\rm *, T})$.  In this case, the source's profit is the optimal value of problem \eqref{Refor}. Combining this and the argument that the optimal value of problem \eqref{Refor} leads to  the maximal achievable profit for the source, we show that the optimal solution $(K^o,\bs{x}^o)$  to problem \eqref{Refor} leads to the optimal time-dependent pricing scheme in \eqref{zzzz}.

\subsection{Proof of Proposition \ref{P222}}\label{ProofP2}

Define
\begin{align}
\widehat{DF}(x,y,z)\triangleq \int_0^x [f(t+y)-f(t+z)]dt.
\end{align}

We first consider the following lemma:
\begin{lemma}\label{L3}
	When the AoI cost function $f(t)$ is convex, the function $\tilde{DF}(\cdot)$ satisfies, for all $y\geq z$,
	\begin{align}
\widehat{DF}(x,y+a,y)\geq \widehat{DF}(x,z+a,z).
\end{align}

\end{lemma}

\begin{IEEEproof}
Consider the derivative of function $\widehat{DF}(x,y+a,y)$.
\begin{align}
\frac{\partial \widehat{DF}(x,y+a,y)}{\partial y}=\int_{0}^xf'(t+y+a)dt-\int_{0}^x f'(t+y)dt.
\end{align}
By the convexity of $f(\cdot)$, we have $f'(t+y+a)\geq f'(t+y)$. Hence, we have
\begin{align}
\int_{0}^xf'(t+y+a)dt&\geq\int_{0}^x f'(t+y)dt\nonumber\\
\Longrightarrow~~~~\frac{\partial \widehat{DF}(x,y+a,y)}{\partial y}&\geq 0,
\end{align}
which indicates that $\widehat{DF}(x,y+a,y)$ is non-decreasing in $y$. Therefore, we complete the proof.
\end{IEEEproof}

In the following, we prove Proposition \ref{P222} by induction, showing that for an arbitrary time-dependent pricing scheme yielding more than $K>1$ updates, there always exists a pricing scheme leading to a single-update equilibrium that is more profitable. The following example illustrates this with a linear AoI cost function.

		\begin{itemize}
	\item \textbf{Base case:} 	For arbitrary update intervals $\bs{x}=\{x_1,x_2,x_3\}$, the destination's payment is	
	\begin{align}
&F(x_1+x_2+x_3)-F(x_1)-F(x_2)-F(x_3)\nonumber\\
	&-\widehat{DF}(x_3,x_1+x_2,x_2)-C(2).
	\end{align}
	Consider a new update intervals $\{\tilde{x}_k\}$ such that 
	\begin{align}
	{x}'_k=\begin{cases}
	x_2,&~{\rm if}~k=1,\\
	x_1+x_3,&~{\rm if}~k=2.
	\end{cases}
	\end{align}
	The destination's payoff becomes
	\begin{align}
	&F(x_1+x_2+x_3)-F(x_1+x_3)-F(x_2)-C(1),\nonumber\\
	=&F(x_1+x_2+x_3)-F(x_1)-F(x_2)-F(x_3)\nonumber\\
	&-\widehat{DF}(x_3,x_1,0)-C(1),\nonumber\\
	\overset{(a)}{\geq}& F(x_1+x_2+x_3)-F(x_1)-F(x_2)-F(x_3)\nonumber\\
	&-\widehat{DF}(x_3,x_1+x_2,x_2)-C(1),
	\end{align}
where $(a)$ is due to Lemma \ref{L3}. This means that for arbitrary pricing scheme leading to two updates, there always exists another pricing scheme with one update that is strictly more profitable.
	
	\item  \textbf{Induction step:} Let $K\geq n$ and suppose that the statement is true for $n=K$,
	the source's profit is 
\begin{align}
&F(x_1+x_2+x_3)-F(x_1)-F(x_2)-F(x_3)\nonumber\\
&-\widehat{DF}(x_3,x_1+x_2,x_2)+\sum_{k=3}^K \widehat{DF}(x_{k+1},x_k,0)-C(K).\label{Eq46}
\end{align}
Consider new update intervals $\bs{x}'_k$ such that 
\begin{align}\label{zzz}
x'_k=\begin{cases}
x_2,&~{\rm if}~k=1,\\
x_1+x_3,&~{\rm if}~k=2,\\
x_{k+1},&~{\rm otherwise}.
\end{cases}
\end{align}
The source's profit becomes
\begin{align}\hspace{-1cm}
~&F(x_1+x_2+x_3)-F(x_1+x_3)-F(x_2)\nonumber\\&+\widehat{DF}(x_4,x_3+x_1,x_1)+\sum_{k=4}^K\widehat{DF}(x_{k+1},x_k,0)\nonumber\\
&-C(K-1),\nonumber\\
\overset{(b)}{\geq}~ 
&F(x_1+x_2+x_3)-F(x_1+x_3)-F(x_2)\nonumber\\
&+\widehat{DF}(x_4,x_3,0)+\sum_{k=4}^K\widehat{DF}(x_{k+1},x_k,0)-C(K-1),\nonumber\\
= ~& F(x_1+x_2+x_3)-F(x_1)-F(x_2)-F(x_3)\nonumber\\
~&-\widehat{DF}(x_3,x_1,0)+\sum_{k=3}^K\widehat{DF}(x_{k+1},x_k,0)-C(K-1),\nonumber\\
\overset{(c)}{\geq} ~& F(x_1+x_2+x_3)-F(x_1)-F(x_2)-F(x_3)\nonumber\\
~&-\widehat{DF}(x_3,x_1+x_2,x_2)+\sum_{k=3}^K\widehat{DF}(x_{k+1},x_k,0)\nonumber\\
&-C(K-1),\label{Eq47}
\end{align}
where $(b)$ and $(c)$ are due to Lemma \ref{L3}. By comparing \eqref{Eq47} and \eqref{Eq46}, we see that adopting \eqref{zzz} strictly increases the source's profit.
 
\end{itemize}
Based on induction, we can show that we can find a $(K'-1)$-udpate policy would be more profitable than the $(K',\bs{x}')$ policy. This eventually leads to the conclusion that a single update policy is the most profitable. 
\subsection{Proof of Corollary \ref{P11}}\label{ProofCoro1}

By Proposition \ref{P222}, the problem in \eqref{Refor} becomes
\begin{align}
&\max_{x\in[0,T]}DF(T-x,x)=\max_{x\in[0,T]}[F(T)-F(x)-F(T-x)].\label{Eq48}
\end{align}	
The necessary condition for optimality yields 
\begin{align}
f(x^{\rm *,T})=f(T-x^{\rm *,T}),\label{Eq49}
\end{align}
which indicates $x^*=T/2$. Proposition \ref{P222} implies that the optimal price at $T/2$ is 
\begin{align}
p(T/2)=DF(T/2,T/2).
\end{align}
Notice that the aggregate AoI cost reduction by updating at $T/2$ is $DF(T/2,T/2)$ and is $DF(T-x,x)$ at any other time instance $x\neq T/2$. In addition, the second-order derivative of $DF(T-x,x)$ with respect to $x$ is 
\begin{align}
\frac{\partial^2 DF(T-x,x)}{\partial x^2}=-f'(x)-f'(T-x)<0,
\end{align}
which implies that $DF(T-x,x)$ is strictly convex in $x$. Therefore, $DF(T-x,x)<DF(T/2,T/2)$ for all other time instances $x\neq T/2$. Under the pricing scheme $p(t)=DF(T/2,T/2)$ for all $t\in\mathcal{T}$, the destination would only update at $T/2$, which completes the proof.

%
\subsection{Proof of Lemma \ref{L4}}\label{ProofL4}

	With $K$ being fixed, the problem in \eqref{D} is reduced to a convex problem. Since it is easy to verify that the Slater's condition is satisfied, the following KKT conditions are both necessary and sufficient for the optimality of the reduced problem:
	\begin{subequations}
		\begin{align}
		F'(x_k)=f(x_k)&=\mu,~~~\forall k\in\mathcal{K}(K^{\rm *,Q}+1),\label{KKT1}\\
		\mu \left(\sum_{k=1}^{K+1} x_k-T\right)&=0, \label{KKT2}
		\end{align}
	\end{subequations}
	where $\mu$ is the dual variable corresponding to constraint \eqref{D2}. Combining \eqref{KKT1} and \eqref{KKT2}, we see that \eqref{L4eq} is an optimal solution to \eqref{KKT1} and \eqref{KKT2}. In addition, since $f(x_k)$ is an increasing function and thus is \textit{one-to-one}. This implies that \eqref{L4eq} is the \textit{unique optimal solution} to \eqref{KKT1} and \eqref{KKT2}. We complete the proof.

\subsection{Proof of Proposition  \ref{T5}}\label{ProofP3}
We first prove \eqref{Ds} by considering the following two cases:
\begin{itemize}
	\item Suppose that there exists an optimal solution to \eqref{Bilevel} such that
	$\sum_{k=1}^{K^{\rm *, Q}}p_k^{*}>F(T)-(K^{\rm *, Q}+1)F(T/(K^{\rm *, Q}+1))$. In this case, the destination's overall cost (the aggregate AoI cost plus the payment) when given 0 update is $\Upsilon(0,{\Pi}_q^{\rm *})=F(T)$ while $\Upsilon(K^{\rm *, Q},{\Pi}_q^{\rm *})=(K^{\rm *, Q}+1)F(T/(K^{\rm *, Q}+1))+\sum_{k=1}^{K^{\rm *, Q}}p_k^{*}$. By the case condition, we have $\Upsilon(K^{\rm *, Q},{\Pi}_q^{\rm *})>\Upsilon(0,{\Pi}_q^{\rm *})$, violating the constraint that $K^{\rm *, Q}$ minimizes $\Upsilon(K,{\Pi}_q^{\rm *})$. Hence, we must have $\sum_{k=1}^{K^{\rm *, Q}}p_k^{\rm *}\leq F(T)-(K^{\rm *, Q}+1)F(T/(K^{\rm *, Q}+1))$.
	\item Suppose that there exists an optimal solution to \eqref{Bilevel} such that
	$\sum_{k=1}^{K^{\rm *, Q}}p_k^{*}<F(T)-(K^{\rm *, Q}+1)F(T/(K^{\rm *, Q}+1))$. In this case, there exists another quantity-based pricing $\tilde{p}_1$ satisfying $\tilde{p}_1>{p}^{\rm *, Q}_1$, $\tilde{p}_k=p_k^{*}$ 
	for all remaining $k>1$, and 
	\begin{align}
	\sum_{k=1}^{K^{\rm *, Q}}\tilde{p}_k\leq F(T)-(K^{\rm *, Q}+1)F\left(\frac{T}{K^{\rm *, Q}+1}\right).\label{DDD}
	\end{align} For the new pricing scheme $
	\tilde{\Pi}_q^{\rm *}$, the overall cost $\Upsilon(K^{\rm *, Q},{\Pi}_q^{\rm *})$ is increased by the same constant $\tilde{p}_1-p^{\rm *}_1$ for all positive $K'$. Combining this with \eqref{DDD}, we see that $K^{\rm *, Q}$ still satisfies the constraint that $K^{\rm *, Q}$ minimizes the overall cost $\Upsilon(K^{\rm *, Q},{\Pi}_q^{\rm *})$ but the pricing scheme $\tilde{\Pi}_q$ increases the source's profit. This implies that $(\tilde{\Pi}_q,K^{\rm *, Q},\bs{x}^{\rm *, Q})$ achieves a higher profit than $({\Pi}_q^{\rm *},K^{\rm *, Q},\bs{x}^{\rm *, Q})$, which contradicts with the fact that $({\Pi}_q^{\rm *},K^{\rm *, Q},\bs{x}^{\rm *, Q})$ is optimal to \eqref{Bilevel}. Hence, we must have $\sum_{k=1}^{K^{\rm *, Q}}p_k^{*}\geq F(T)-(K^{\rm *, Q}+1)F(T/(K^{\rm *, Q}+1))$.
	\end{itemize}
Combining the above discussions, we complete the proof for \eqref{Ds}. To ensure that $K^*$ satisfies constraint \eqref{C3}, we see that $K^*$ must satisfy \eqref{Da}. We complete the proof.

\subsection{Proof of Proposition \ref{P3}}\label{ProofP4}
Let $K^{\rm *, Q}$ be the optimal solution to problem \eqref{Bilevel}. 
 By the definition of $\hat{K}$ in \eqref{hatK}, we have $\hat{K}\leq K^*\leq \hat{K}+1$. 
 The convexity of the objective in \eqref{eq28} implies that the objective of \eqref{eq28} (which is also the objective of problem \eqref{PBK}) is non-decreasing in $K$ for all $K\leq K^*$ and is non-increasing in $K$ for all $K\geq K^{\rm *, Q}$. 
 This implies an optimal solution to problem \eqref{PBK} is either $\hat{K}$ or $\hat{K}+1$. 

	
\subsection{Proof of Theorem \ref{P2}}\label{ProofT1}
The social cost minimization problem in \eqref{SCM-F} is
\begin{align}
&\min_{K\in\mathbb{N},\bs{x}\in\mathcal{X}(K)}~\sum_{k=1}^{K+1}F(x_k)+C(K),\nonumber\\
\overset{(a)}{=}&~~~~~\min_{K\in\mathbb{N}}~(K+1)F(T/(K+1))+C(K),\label{4z}
\end{align}
where $\mathcal{X}(K)\triangleq\{\bs{x}: \bs{x}\in\mathbb{R}_{++}^{K+1},~\sum_{k=1}^{K+1}x_k=T\}$; (a) is due to the similar reason as in the proof of Lemma \ref{L4}. Specially, if we fix $K$, it is readily verified that the reduced problem (of optimizing $\bs{x}$) satisfies the Slater's condition. Hence, the following KKT conditions are both necessary and sufficient for the optimality of the reduced problem:
\begin{subequations}
	\begin{align}
	F'(x_k)=f(x_k)&=\lambda,~~~\forall k\in\mathcal{K}(K+1),\\
	\lambda \left(\sum_{k=1}^{K+1} x_k-T\right)&=0, 
	\end{align}
\end{subequations}
where $\lambda$ is the dual variable corresponding to constraint \eqref{D2}. By Definition \ref{Def7} and Proposition \ref{P3}, we see that the optimal quantity-based pricing $\Pi^*_q$ is surplus extracting. From Lemma \ref{L1}, the optimal quantity-based pricing scheme is optimal among all possible pricing schemes.


\subsection{Proof of Lemma \ref{L5}}\label{ProofL5}
We define the value function at time $S_k$ as the maximal objective value of the SCM-U Problem in \eqref{D-SCM}, satisfying 
\begin{align}\label{ProofL5P}
  V_{c,k}\triangleq  &\min_{\mathcal{S}_{-k}}\quad 
\left[\delta^{S_{k-1}^o} F_{\delta}(S_k-S_{k-1})\right.\nonumber\\
&~~~~~~~\left.+\lim_{K\rightarrow \infty} \sum_{k'=k}^{K}\delta^{S_{k'}} \left[F_{\delta}(S_{k'+1}-S_{k'})+c\left(\mathcal{S}\right)\right]\right]\\
	&{\rm s.t.}\quad S_{k'}\geq S_{k'-1},~\forall k'\geq k,
\end{align}
where $\mathcal{S}_{-k}=\{S_{j}\}_{j\geq k}$.
From the objective value of the SCM-U Problem in \eqref{D-SCM}, we observe that such value functions are related based on the following equation:
\begin{align}
   	&V_{c,k}=\nonumber\\
   	&\min_{S_k}\left[F_\delta(S_k-S_{k-1})+\delta^{S_k-S_{k-1}} c(S_k-S_{k-1})\right.\nonumber\\
   	&\left.+\delta^{S_k-S_{k-1}} V_{c,k+1}\right],~{\rm s.t.}~S_k\geq S_{k-1}. \label{ahahds}
\end{align}
We define $\tilde{\mathcal{S}}=\{\tilde{S}_k\}_{k\in\mathbb{N}}$ such that $\tilde{S}_{j}=S_{j+k}$ for all $j\in\mathbb{N}$. The problem in \eqref{ProofL5P} becomes
	\begin{align}
	\min_{\tilde{\mathcal{S}}\in\Phi}\quad &
    	F_{\delta}(\tilde{S}_1)+\lim_{K\rightarrow \infty} \sum_{k=1}^{K}\delta^{\tilde{S}_{k}} \left[F_{\delta}(\tilde{S}_{k+1}-\tilde{S}_{k})+c\left(\bar{x}(\tilde{\mathcal{S}})\right)\right],
	\end{align}
which is equivalent to the problem in \eqref{D-SCM}. This implies that in fact $V_{c,k}=V_{c,j}$ for all $j,k\in\mathbb{N}$. Hence, replacing $V_{c,k}$ and $V_{c,k+1}$ by $V_{c}$ in \eqref{ahahds} proved Lemma \ref{L5}.

	\subsection{Proof of Proposition \ref{P5}}\label{ProofP5}
Lemma \ref{L5} implies that
\begin{align}
    V_c&=\frac{F_\delta(x^o)+\delta^{x^o}c(x^o)}{1-\delta^{x^o}}, \label{Proof-Eq1}\\
    0&= f(x^o)+\ln(\delta) c(x^o)+ c'(x^o)+\ln(\delta)  V_c. \label{Proof-Eq2}
\end{align}
Combining \eqref{Proof-Eq1} and \eqref{Proof-Eq2}, we have
\begin{align}
    &~F_\delta(x^o)-\frac{1}{\ln(\delta^{-1})}(1-\delta^x)f(x)=(1-\delta^{x^o})c'(x^o)-c(x^o)\nonumber\\
   \overset{(a)}{\Longrightarrow}&~\frac{\int_0^{x^o}(1-\delta^t)f'(t)dt}{\ln(\delta^{-1})} =(1-\delta^{x^o})c'(x^o)-c(x^o),\nonumber\\
    \Longrightarrow&\int_0^{x^o}(1-\delta^t)f'(t)dt =\ln(\delta^{-1})[c(x^o)-(1-\delta^{x^o})c'(x^o)],\nonumber\\
    \overset{(b)}{\Longrightarrow}&~\int_0^{x^o}(1-\delta^t)f'(t)dt =\ln(\delta^{-1})\nonumber\\
    &\times \left[c(x^o)-\int_{0}^{x^o}\left(\ln(\delta)\delta^t c'(t)+(1-\delta^t) c''(t)\right)dt
    \right],
\end{align}
where (a) and (b) are due to the fact that $\int g(t)h'(t)dt=g(x)h(x)-\int g'(t)h(t)dt$ for all differentiable functions $g(x),h(x)$.

	\subsection{Proof of Lemma \ref{LLL2}}\label{ProofL555}
By the individual rationality constraint in \eqref{IRU}, we have
\begin{align}
P_\delta(\mathcal{S}^o(\Pi),\Pi)\leq F_\delta(\infty)-\Gamma_\delta(\mathcal{S}^o(\Pi)).
\end{align}
Hence, the source's profit thus is
\begin{align}
&P_\delta(\mathcal{S}^o(\Pi))-C_\delta(K^o(\Pi))\nonumber\\
\leq &~F_\delta(\infty)-\Gamma_\delta(\mathcal{S}^*(\Pi))-C_\delta(\mathcal{S}^*(\Pi)),\nonumber\\
\leq &~F_\delta(\infty)-\min_{\mathcal{S}\in\Phi}[\Gamma_\delta(\mathcal{S})+C_\delta(\mathcal{S})]. \label{eq49999}
\end{align}
Hence, if the profit of a pricing scheme attains the upper bound in \eqref{eq49999}, it achieves the maximal profit among all pricing schemes.

\subsection{Proof of Proposition \ref{P8}}\label{ProofP7}
First, we prove $\mathcal{L}(x,Q)$ is Lipschitz continuous and it is readily verified that the corresponding Lipschitz continuous $L_{\mathcal{L}}$ is given by
$L_{\mathcal{L}}\triangleq 4\max_{t\geq 0}[\delta^tf(t)]+L_c+Q \ln(\delta^{-1})$. That is,
\begin{align}
    |L(x_1,Q)-L(x_2,Q)|\leq L_{\mathcal{L}}|x_1-x_2|,~\forall x_1,x_2\geq 0.
\end{align}

Second, to show the optimal solution $x^\star_t\leq \tilde{x}(Q)$, we substitute $\bar{x}$ into
the objective in \eqref{FP}, which yields
\begin{align}
    \mathcal{L}(\bar{x},Q)=-Q(1-\delta^x). \label{Suboptimal1}
\end{align} 
Note that
\begin{align}
    \mathcal{L}(x,Q)\leq &~\delta^x \int_{0}^x\delta^tf(x+t)dt -Q(1-\delta^x)\nonumber\\
    \overset{(a)}{<} &~\frac{\delta^x\zeta^x}{\ln((\delta\zeta)^{-1})}-Q(1-\delta^x)\nonumber\\
    \leq & ~[\delta\cdot\max(1,\zeta)]^x\left(\frac{1}{\ln((\delta\zeta)^{-1})}+Q\right)-Q \label{Suboptimal2}.
\end{align}
where $(a)$ is due to Assumption \ref{Assum2}.
Combining \eqref{tildex}, \eqref{Suboptimal1}, and \eqref{Suboptimal2}, we have
\begin{align}
    L({x},Q)\leq L(\bar{x},Q)\leq \max_{x'\geq 0}L(x',Q),~\forall {x}\geq \tilde{x}(Q),
\end{align}
which implies that, when $Q$ is fixed, the optimal solution  of $L(x,Q)$  must live on $[0,\tilde{x}(Q)]$. 

\subsection{Proof of Proposition \ref{Asy}}\label{ProofP8}
Let $\zeta_{\delta,t}$ and $\zeta_\delta^*$ be the source's discounted profits under the equal spacing time-dependent pricing and a surplus-extracting pricing, respectively.
Let $x^o$ be the socially optimal interarrival time. We have
	\begin{align}
	    \frac{\zeta_{\delta,t}}{\zeta_\delta^*}&\geq\frac{F_\delta(2x^o)-(1+\delta^{x^o})F_\delta(x^o)-\delta^{x^o}c(x^o)}{(1-\delta^{x^o})F_\delta(\infty)-F_\delta(x^o)-c(x^o)}\nonumber\\
	    &\geq \frac{F_\delta(x^o)-(1+\delta^{x^o})F_\delta(x^o)-\delta^{x^o}c(x^o)}{(1-\delta^{x^o})F_\delta(\infty)-F_\delta(x^o)-c(x^o)}\label{Eq84}.
	\end{align}
	We start with the following lemmas:
\begin{lemma}\label{L12}
    If $\lim_{\delta\rightarrow 0}\frac{A(\delta)}{B(\delta)}\rightarrow 1$ and $\lim_{\delta\rightarrow 0}\frac{C(\delta)}{D(\delta)}\rightarrow 1$, then $\lim_{\delta\rightarrow 0}\frac{A(\delta)+C(\delta)}{B(\delta)+D(\delta)}\rightarrow 1$.
\end{lemma}	
\begin{IEEEproof}
  To see this, we start with the following equation
	\begin{align}
	    \frac{A(\delta)}{B(\delta)}=\frac{A(\delta)\left(1+\frac{D(\delta)}{B(\delta)}\right)}{B(\delta)+D(\delta)}=\frac{A(\delta)+\frac{A(\delta)}{B(\delta)}\frac{D(\delta)}{C(\delta)}C(\delta)}{B(\delta)+D(\delta)}. \label{Eq85}
	\end{align}
	From \eqref{Eq85}, we have
	\begin{align}
	    \lim_{\delta\rightarrow 0}\frac{A(\delta)}{B(\delta)}&=\lim_{\delta\rightarrow 0}\frac{A(\delta)+\frac{A(\delta)}{B(\delta)}\frac{D(\delta)}{C(\delta)}C(\delta)}{B(\delta)+D(\delta)}\nonumber\\
	    &=\lim_{\delta\rightarrow 0}\frac{A(\delta)+C(\delta)}{B(\delta)+D(\delta)}=1.
	\end{align}
\end{IEEEproof}

	\begin{lemma}\label{L13}
	The following equation holds:
	    $\lim_{\delta \rightarrow 0 }\frac{\int_{0}^{x^o}\delta^tf(t)dt}{\int_{0}^{\infty}\delta^tf(t)dt}=1$.
	\end{lemma}
	\begin{IEEEproof}
	We have
		\begin{align}
	    &\lim_{\delta \rightarrow 0 }\frac{\int_{0}^{x^o}\delta^tf(t)dt}{\int_{0}^{\infty}\delta^tf(t)dt}\nonumber\\
	    =&\lim_{\delta \rightarrow 0 }\frac{\int_{0}^{x^o}\delta^tf(t)dt}{\sum_{k=0}^{\infty}\delta^{kx^o}\int_{0}^{x^o}\delta^tf(t+(k-1)x^o)dt}\nonumber\\
	    \overset{(a)}{\geq} &\lim_{\delta \rightarrow 0 }\frac{\delta^{\hat{x}}f(\hat{x})x^o}{\delta^{\hat{x}}f(\hat{x})x^o+\delta^{x^o}\sum_{k=1}^{\infty}(\delta\zeta)^{(k-1)x^o}\int_{0}^{x^o}A(\delta\zeta)^tdt}\nonumber\\
        \overset{(b)}{\geq} &\lim_{\delta \rightarrow 0 }\frac{f(\hat{x})x^o}{f(\hat{x})x^o+\delta^{x^o-\hat{x}}A\int_{0}^{x^o}\gamma^{t-x^o}dt}=1,
	\end{align}
		where (a) is due to Assumption \ref{Assum2} that $f(t)\leq A\zeta^t$ and due to the mean value theorem, there exists a $\hat{x}\in(0,x^o)$ such that $\delta^{\hat{x}}f(\hat{x})x^o=\int_{0}^{x^o}\delta^tf(t)dt$; (b) is due to that $\zeta\delta \leq \gamma$ according to Assumption \ref{Assum2}.
	\end{IEEEproof}

From Lemma \ref{L13}, we have $\lim_{\delta\rightarrow 0}\frac{F_\delta(x^o)}{(1-\delta^{x^o})F_\delta(\infty)}=1$. From Lemma \ref{L12} and \eqref{Eq84}, 
We further have that
	\begin{align}
	    1 \geq \lim_{\delta\rightarrow 0}\frac{\zeta_{\delta,t}}{\zeta_\delta^*}&\geq\lim_{\delta\rightarrow 0}\frac{F_\delta(x^o)-(1+\delta^{x^o})F_\delta(x^o)-\delta^{x^o}c(x^o)}{(1-\delta^{x^o})F_\delta(\infty)-F_\delta(x^o)-c(x^o)}
	    =1.
	\end{align}

\subsection{Proof of Lemma \ref{L66}}\label{ProofL7}
Taking the derivative of the destination's objective in \eqref{destination-2} with respect to $S_k$ yields 
\begin{align}
&\delta^{S_k} \ln(\delta) [p_k+F_\delta(S_{k+1}-S_{k})]+\delta^{S_{k-1}}f_\delta(S_k-S_{k-1})+  \nonumber\\
&-\delta^{S_k}f_\delta(S_{k+1}-S_k)=0,~\forall k\in\mathbb{N},\\
&\ln(\delta) p_k+f(S_k-S_{k-1})+\ln (\delta) F_\delta(S_{k+1}-S_{k})\nonumber\\
=&f_\delta(S_{k+1}-S_k),~\forall k\in\mathbb{N},
\label{Z1}
\end{align}
For each $k\in\mathbb{N}$, summing \eqref{Z1} over all $j\geq k$ leads to Lemma \ref{L66}.

\subsection{Proof of Lemma \ref{L8}}\label{ProofL8}

The problem in \eqref{PP} is equivalent to
	\begin{align}
	&\max_{\mathcal{S}\in\Phi}~ \frac{1}{\ln(\delta^{-1})} f_\delta(S_1)-\lim_{K\rightarrow \infty} \sum_{k=1}^{K}\delta^{S_{k}} \left[
	F_{\delta}({S}_{k+1}-{S}_{k})+c(\bar{x})\right]\nonumber\\
\overset{(a)}{=}&\max_{S_1,~\tilde{\mathcal{S}}\in\Phi}~ \frac{1}{\ln(\delta^{-1})} f_\delta(S_1)-\delta^{S_1}  \min_{\tilde{\mathcal{S}}\in\Phi} \left[\lim_{K\rightarrow \infty} \sum_{k=0}^{K}\delta^{\tilde{S}_{k}} \right.\nonumber\\
&\left.\times\left[
	F_{\delta}(\tilde{S}_{k+1}
	-\tilde{S}_{k})+c(\bar{x}))\right]\right],\nonumber\\
\overset{(b)}{=}&\max_{S_1,~\tilde{\mathcal{S}}\in\Phi}~ \frac{1}{\ln(\delta^{-1})} f_\delta(S_1)-\delta^{S_1} (V_c+c(x^o)),
	\end{align}
where  $V_c$ is the minimal discounted social cost $V_c$ derived in \eqref{MSC}. In (a), we replace $\{S_k\}_{k\geq 2}$ by $\tilde{\mathcal{S}}$ such that $\tilde{S}_k=S_k-S_1$ for all $k\in\mathcal{K}$; we derive (b) based on similarity between the structure and the SCM-U Problem in \eqref{D-SCM}.

\end{document}